\documentclass[paper,notoc]{JHEP3}
\usepackage{epsfig,cite}
\usepackage{amsbsy} 
\usepackage{epsfig}
\usepackage{graphicx}

\newcommand\new{\newcommand}         

\newcommand{\mh}{\ensuremath{m_\mathrm{H}}}

\def\beq{\begin{equation}}   
\def\eeq{\end{equation}}
\def\bea{\begin{eqnarray}}  
\def\eea{\end{eqnarray}}

\def\pthard{p_{\rm T}^{\rm hard}}
\def\ptsoft{p_{\rm T}^{\rm soft}}

\def\ln{\hbox{ln}}

\new{\emem}{{\ifmmode\mathrm{e}^-\else e$^-$\fi}}
\new{\epem}{{\ifmmode\mathrm{e}^+\else e$^+$\fi}}
\new{\zbo}  {{\ifmmode\mathrm{Z}\else Z\fi}}
\new{\wpm} {{\ifmmode\mathrm{W}^\pm\else W$^\pm$\fi}}
\new{\wbo} {{\ifmmode\mathrm{W}\else W\fi}}
\new{\epm} {{\ifmmode\mathrm{e^+e^-}\else $\mathrm{e^+e^-}$\fi}}
\new{\qq}  {{\ifmmode\mathrm{q}\else q\fi}}
\new{\qqb} {{\ifmmode\bar{\mathrm{q}}\else $\bar{\mathrm{q}}$\fi}}
\new{\tq}  {{\ifmmode\mathrm{t}\else t\fi}}
\new{\tqb} {{\ifmmode\bar{\mathrm{t}}\else $\bar{\mathrm{t}}$\fi}}
\new{\bq}  {{\ifmmode\mathrm{b}\else b\fi}}
\new{\bqb} {{\ifmmode\bar{\mathrm{b}}\else $\bar{\mathrm{b}}$\fi}}
\new{\ttbar}{\tq\tqb}
\new{\qqbar}{\qq\qqb}
\new{\gu}  {{\ifmmode\mathrm{g}\else g\fi}}
\new{\qqbarg}{\qq\qqb\gu}
\new{\pp}  {{\ifmmode\mathrm{p}\else p\fi}}
\new{\K}{\ensuremath{K}}
\new{\hh}  {{\ifmmode h\else $h$\fi}}
\new{\HH}  {{\ifmmode \mathrm{H}\else $\mathrm{H}$\fi}}
\new{\higgs}{\ensuremath{\mathrm{H}}}
\new{\W}{\ensuremath{\mathrm{W}}}
\new{\WW}{\W\W}
\new{\mll}{\ensuremath{m_{\ell\ell}}}
\new{\bbbar}{\ensuremath{b\bar{b}}}
\new{\ppbar}{\ensuremath{\mathrm{p}\bar{\mathrm{p}}}}
\new{\muf}{\ensuremath{\mu_\mathrm{F}}}
\new{\mur}{\ensuremath{\mu_\mathrm{R}}}
\new{\fe}  {{\ifmmode f\else $f$\fi}}
\new{\lp}  {{\ifmmode \ell\else $\ell$\fi}}
\new{\XX}  {{\ifmmode X\else $X$\fi}}
\new{\Vp}  {{\ifmmode V\else $V$\fi}}
\new{\Kzs} {{\ifmmode\mathrm{K}_\mathrm{S}^0\else $\mathrm{K}_\mathrm{S}^0$\fi}}
\new{\Kzl} {{\ifmmode\mathrm{K}_\mathrm{L}^0\else $\mathrm{K}_\mathrm{L}^0$\fi}}
\new{\Kp} {{\ifmmode\mathrm{K}\else $\mathrm{K}$\fi}}
\new{\ppHWW} {{\ifmmode\pp\pp\rightarrow\HH\rightarrow\wbo\wbo
                             \else $\pp\pp\rightarrow\HH\rightarrow\wbo\wbo$\fi}}
\new{\ppHWWlept} {{\ifmmode\pp\pp\rightarrow\HH\rightarrow\wbo\wbo\rightarrow\lp\nu\lp\nu
                             \else $\pp\pp\rightarrow\HH\rightarrow\wbo\wbo\rightarrow\lp\nu\lp\nu$\fi}}
\new{\HWWlept} {{\ifmmode\HH\rightarrow\wbo\wbo\rightarrow\lp\nu\lp\nu
                             \else $\HH\rightarrow\wbo\wbo\rightarrow\lp\nu\lp\nu$\fi}}

\new{\LEP}        {\mbox{\small\textsc{LEP}}}
\new{\LEPONE}     {\mbox{\small\textsc{LEP1}}}
\new{\LEPTWO}     {\mbox{\small\textsc{LEP2}}}
\new{\CERN}       {\mbox{\small\textsc{CERN}}}
\new{\ALEPH}      {\mbox{\small\textsc{ALEPH}}}
\new{\DELPHI}     {\mbox{\small\textsc{DELPHI}}}
\new{\LD}         {\mbox{\small\textsc{L3}}}
\new{\OPAL}       {\mbox{\small\textsc{OPAL}}}
\new{\SPS}        {\mbox{\small\textsc{SPS}}}
\new{\TEVATRON}   {\mbox{\small\textsc{TEVATRON}}}
\new{\LHC}        {\mbox{\small\textsc{LHC}}}
\new{\FERMILAB}   {\mbox{\small\textsc{FERMILAB}}}
\new{\CDF}        {\mbox{\small\textsc{CDF}}}
\new{\DZERO}      {\mbox{\small\textsc{D0}}}
\new{\CTEQ}        {\mbox{\small\textsc{CTEQ}}}
\new{\FNAL}        {\mbox{\small\textsc{FNAL}}}
\new{\ATLAS}        {\mbox{\small\textsc{ATLAS}}}
\new{\CMS}        {\mbox{\small\textsc{CMS}}}

\new{\eV}         {{\ifmmode {\mathrm{ eV}}\else ${\mathrm{ eV}}$\fi}}
\new{\MeV}        {{\ifmmode {\mathrm{ MeV}}\else ${\mathrm{ MeV}}$\fi}}
\new{\MeVc}       {{\ifmmode {\mathrm{ MeV}}/c\else ${\mathrm{ MeV}}/c$\fi}}
\new{\MeVcc}      {{\ifmmode {\mathrm{ MeV}}/c^2\else ${\mathrm{ MeV}}/c^2$\fi}}
\new{\GeV}        {{\ifmmode {\mathrm{ GeV}}\else ${\mathrm{ GeV}}$\fi}}
\new{\GeVc}       {{\ifmmode {\mathrm{ GeV}}/c\else ${\mathrm{GeV}}/c$\fi}}
\new{\GeVcc}      {{\ifmmode {\mathrm{ GeV}}/c^2\else ${\mathrm{GeV}}/c^2$\fi}}
\new{\TeV}        {{\ifmmode {\mathrm{ TeV}}\else ${\mathrm{ TeV}}$\fi}}

\new{\fb}        {{\ifmmode {\mathrm{ fb}}\else ${\mathrm{ fb}}$\fi}}
\new{\fbinv}   {{\ifmmode {\mathrm{ fb}^{-1}}\else ${\mathrm{ fb}^{-1}}$\fi}}
\new{\pb}        {{\ifmmode {\mathrm{ pb}}\else ${\mathrm{ pb}}$\fi}}
\new{\pbinv}   {{\ifmmode {\mathrm{ pb}^{-1}}\else ${\mathrm{ pb}^{-1}}$\fi}}

\new{\FEHiP}        {\mbox{\small\textsc{FEHiP}}}

\new{\Mz}         {{\ifmmode M_{\mathrm{ Z}}
                    \else $M_{\mathrm{ Z}}$\fi}}
\new{\Mzsq}       {{\ifmmode M^2_{\mathrm{ Z}}
                    \else $M^2_{\mathrm{ Z}}$\fi}}
\new{\Mw}         {{\ifmmode M_{\mathrm{ W}}
                    \else $M_{\mathrm{ W}}$\fi}}
                    
\new{\MH}         {{\ifmmode M_{\mathrm{ H}}
                    \else $M_{\mathrm{ H}}$\fi}}

\new{\as}[1]      {{\ifmmode\alpha^{#1}_s
                    \else$\alpha^{#1}_s$\fi}}
\new{\asx}[1]      {{\ifmmode a^{#1}_s
                    \else $a^{#1}_s$\fi}}
\new{\asb}[1]     {{\ifmmode\overline{\alpha}^{#1}_s
                    \else $\overline{\alpha}^{#1}_s$\fi}}
\new{\asmz}       {{\ifmmode\alpha_s(\Mzsq)
                    \else $\alpha_s(\Mzsq)$\fi}}
\new{\lqcd}       {{\ifmmode\Lambda_{\mathrm{ QCD}}
                    \else $\Lambda_{\mathrm{ QCD}}$\fi}}
\new{\lqcdsq}     {{\ifmmode\Lambda^2_{\mathrm{ QCD}}
                    \else $\Lambda^2_{\mathrm{ QCD}}$\fi}}
\new{\llla}       {{\ifmmode\Lambda_{\mathrm{ LLA}}
                    \else $\Lambda_{\mathrm{ LLA}}$\fi}} 
\new{\lmsbar}[1]  {{\ifmmode \Lambda^{(#1)}_{\overline{\mathrm{MS}}}
                    \else $\Lambda^{(#1)}_{\overline{\mathrm{MS}}}$\fi}}
\new{\lmsb}       {{\ifmmode \Lambda_{\overline{\mathrm{MS}}}
                    \else $\Lambda_{\overline{\mathrm{MS}}}$\fi}}
\new{\lmsbsq}     {{\ifmmode \Lambda^{2}_{\overline{\mathrm{MS}}}
                    \else $\Lambda^{2}_{\overline{\mathrm{MS}}}$\fi}}
\new{\pt}       {{\ifmmode p_{\mathrm{T}}
                    \else $p_{\mathrm{T}}$\fi}}
\new{\Etmiss}       {{\ifmmode E_{\mathrm{T}}^{\mathrm{miss}}
                    \else $E_{\mathrm{T}}^{\mathrm{miss}}$\fi}}
\new {\MET}{\ensuremath{E_\mathrm{T,miss}}}
\new {\METcut}{\ensuremath{E_\mathrm{T,miss}^{\mathrm{\;cut}}}}
\new {\ET}{\ensuremath{E_\mathrm{T}}}
\new {\kt}{\ensuremath{k_\mathrm{T}}}
\new{\ptlmin}       {{\ifmmode p_{\mathrm{T}}^{\lp\mathrm{min}}
                    \else $p_{\mathrm{T}}^{\lp\mathrm{min}}$\fi}}
\new{\ptlmax}       {{\ifmmode \mathrm{p}_{\mathrm{t,max}}^{\mathrm{cut}}
                    \else $\mathrm{p}_{\mathrm{t,max}}^{\mathrm{cut}}$\fi}} 
\new{\ptlep}       {{\ifmmode p_{\mathrm{T}}^{\mathrm{lepton}}
                    \else $p_{\mathrm{T}}^{\mathrm{lepton}}$\fi}}  
\new{\ptjet}       {{\ifmmode p_{\mathrm{T}}^{\mathrm{jet}}
                    \else $p_{\mathrm{T}}^{\mathrm{jet}}$\fi}}                                      
\new{\ptveto}       {{\ifmmode p_{\mathrm{T}}^{\mathrm{veto}}
                    \else $p_{\mathrm{T}}^{\mathrm{veto}}$\fi}}

\newcommand{\accsigma}{\ensuremath{\sigma_{\mathrm{acc}}}}
\newcommand{\inclsigma}{\ensuremath{\sigma_{\mathrm{incl}}}}

\newcommand{\phill}{\ensuremath{\phi_{\ell\ell}}}

\new{\HERWIG}         {\mbox{\small\textsc{HERWIG}}}
\new{\PYTHIA}         {\mbox{\small\textsc{PYTHIA}}}
\new{\JIMMY}         {\mbox{\small\textsc{JIMMY}}}
\new{\SHERPA}         {\mbox{\small\textsc{SHERPA}}}

\new{\fehip}        {\mbox{\small\textsc{FEHiP}}}
\new{\fewz}        {\mbox{\small\textsc{FEWZ}}}
\new{\hqt}        {\mbox{\small\textsc{HqT}}}
\new{\mcnlo}        {\mbox{\small\textsc{MC@NLO}}}
\new{\pvegas}        {\mbox{\small\textsc{PVEGAS}}}

\title{\boldmath Perturbative QCD effects and the search for a \HWWlept\ signal at the Tevatron}

\author{Charalampos Anastasiou\\
  Institute for Theoretical Physics, ETH Zurich,
  8093 Zurich, Switzerland\\
  E-mail: \email{babis@cern.ch}}
\author{G\"unther Dissertori\\
  Institute for Particle Physics, ETH Zurich,
  8093 Zurich, Switzerland\\
  E-mail: \email{dissertori@phys.ethz.ch}}
\author{Massimiliano Grazzini\\
  INFN, Sezione di Firenze and Dipartimento di Fisica, Universit\`a di Firenze,\\
  I-50019 Sesto Fiorentino, Florence, Italy\\
  E-mail: \email{grazzini@fi.infn.it}}
\author{Fabian St\"ockli\\
  Institute for Particle Physics, ETH Zurich,
  8093 Zurich, Switzerland\\
  E-mail: \email{fabstoec@phys.ethz.ch}}
\author{Bryan Webber\\
  Cavendish Laboratory, Cambridge CB3 0HE, U.K. and\\
  Physics Department, CERN, 1211 Geneva 23, Switzerland\\
  E-mail: \email{webber@hep.phy.cam.ac.uk}}

\abstract{ 
The Tevatron experiments have recently excluded
a Standard Model Higgs boson in the mass range 160 GeV $<\mh <$ 170 GeV
at the $95\%$ confidence level. This
result is based on sophisticated analyses designed to maximize the ratio of accepted
signal to background. In this paper we study the production of a Higgs boson
of mass $\mh=160$ GeV in the $gg \to \higgs \to \WW\to l\nu l\nu$ channel.
We choose a set of cuts like those adopted in the experimental analysis
and compare kinematical distributions of the final state leptons
computed in NNLO QCD to lower-order calculations and to those obtained with the event
generators \PYTHIA, \HERWIG\ and \mcnlo.  We also show that the distribution of the output
from an Artificial Neural Network obtained with the different tools does not show significant differences.
However, the final acceptance computed with \PYTHIA\ is
smaller than those obtained at NNLO and with \HERWIG\ and \mcnlo.
We also investigate the impact of the underlying event and
hadronization on our results. 
} 
\preprint{{ETHZ-IPP-2009-08}, CERN-PH-TH-2009-065, Cavendish-HEP-09/08}

\begin{document}

\section{Introduction}
\label{sec:intro}

Clarifying the role of Higgs bosons in the breaking of electroweak symmetry 
is of  paramount  importance  in improving our understanding of  elementary particle  interactions. The discovery 
of the Standard  Model Higgs boson, or its equivalent in theories  beyond the Standard Model, is a principal objective 
of the high-energy collider experimental program.   

Now is a very exciting time in the Higgs search since the Tevatron experiments have 
become sensitive to potential signals from a Higgs boson with 
production cross-sections of roughly the magnitude predicted 
by the Standard Model (SM).  Both CDF~\cite{Aaltonen:2008ec,CDFnote} 
and D\O ~\cite{DOnote}  have presented  studies  where a  cross-section about $1.5-1.7$ times the SM prediction 
for a Higgs  boson with a mass about twice the W-mass can be  excluded  by the two experiments independently 
with a confidence level of $95\%$.  Recent preliminary combinations ~\cite{Bernardi:2008ee,Phenomena:2009pt}
of the two experiments exclude a SM Higgs boson  in the mass 
range $160- 170\,\GeV$ with a $95\%$ confidence level.  
In this mass range, a signal is predominantly produced in the SM via the 
gluon fusion process $\ppbar \to \higgs \to \WW \to  l l \nu \nu$. 

Higher order calculations for the background  processes  and  especially the 
signal cross-section are indispensable for the study or exclusion of high mass Higgs  bosons at the Tevatron.  
The magnitude  of QCD  corrections for the dominant signal process is extraordinarily large; 
the inclusive cross-section at next-to-next-to-leading order (NNLO) for gluon fusion 
is about three times larger than the  leading order (LO) cross-section. 
This is an even larger \K-factor than that obtained at LHC  energies. 
The Tevatron experiments are sensitive to a Standard Model Higgs boson signal in the mass  
region where gluon fusion is dominant, due to this  particularly large \K-factor and   the small 
theoretical  uncertainty which is attained with calculations at higher orders in perturbation theory.  

Next-to-leading order (NLO) corrections  have been computed in~\cite{Dawson:1990zj,Djouadi:1991tka,Spira:1995rr}, 
where the  anomalously large higher order effects in gluon fusion were first shown. The NNLO corrections 
for the inclusive cross-section have been computed in~\cite{Harlander:2002wh,Anastasiou:2002yz,Ravindran:2003um}. 
The NNLO computation has been consistently improved by resumming the soft-gluon
contributions up to next-to-next-to-leading logarithmic (NNLL) accuracy \cite{Catani:2003zt}.
The result is an additional increase of the cross section that amounts to about $13\%$ at the Tevatron.
The NNLL result is nicely confirmed by the computation of additional soft terms
to N$^3$LO~\cite{Moch:2005ky,Laenen:2005uz,Idilbi:2005ni}
(see also \cite{Ravindran:2005vv,Ravindran:2006cg,Ravindran:2006bu}).
A calculation based on a somewhat different approach has been presented in Ref.~\cite{Ahrens:2008nc}.

The calculations of threshold  effects  have  
provided an invaluable argument  that  by now the bulk of the higher order corrections is accounted for. 
Smaller theoretical effects can also be important  for  setting a precise  exclusion limit on the 
Higgs  boson cross-section at the Tevatron. 
Two-loop electroweak corrections (of about $\sim 5\%$) for gluon fusion Higgs production 
have been computed  in~\cite{Degrassi:2004mx,Aglietti:2004nj,Aglietti:2006yd} , and the full two-loop 
amplitude has been presented in ~\cite{Actis:2008ug,Actis:2008ts}.  
Mixed QCD and Electroweak corrections  have  been computed in~\cite{Anastasiou:2008tj}.  
Recent predictions for the  inclusive cross-section taking into account these effects have been 
presented in~\cite{Anastasiou:2008tj,deFlorian:2009hc}; these theoretical results for the inclusive 
gluon fusion cross-section have been used in setting the exclusion limits of Ref.~\cite{Phenomena:2009pt}.

Electroweak corrections of ${\cal O}(\alpha)$ for the decay \HWWlept\ have  been computed in~\cite{Bredenstein:2006rh}. These reach $6\%$ for the  partial decay width of  the  decay of 
a Higgs  boson to four leptons with mass  of $~170\,\GeV$.  In the region just above  $\mh > 2 M_{\rm W}$, 
which is relevant for the recent Tevatron exclusion limits, Ref.~\cite{Bredenstein:2006rh} reported important differences with the prediction of the program HDECAY~\cite{hdecay}.
These corrections  have not yet been included  into estimates of the Higgs branching ratio 
to leptons,  
and they are not considered in the experimental analysis of Refs.~\cite{Bernardi:2008ee,Phenomena:2009pt}.

The extraction of limits on the Higgs  boson cross-section  from the study of the 
\HWWlept\ process requires a sophisticated analysis. Background processes,  such as 
direct $WW, WZ, ZZ$ production as well as $t \bar{t},W\gamma$ and $W+$ (multi)jet production, 
are dominant. Experimental cuts can suppress  background and enhance the signal
by vetoing hadronic radiation and by  exploiting characteristic  differences in 
lepton angular distributions as well as the large missing transverse momentum in signal events~\cite{Dittmar:1996ss}.  

At the Tevatron, a cut-based analysis alone is not sufficient.  Additional methods 
that exploit efficiently the kinematic features of signal and  background  processes in their finest detail are 
required.   CDF~\cite{CDFnote} and  D\O~\cite{DOnote}  apply cuts on missing  energy and jet activity and  impose  
lepton isolation criteria only for purposes of   a first rough selection which biases the  data samples  towards  
signal events.   After this first cut-based selection, background processes remain dominant and processing of real 
data and Monte-Carlo simulations  with Artificial Neural Network  (ANN) methods follows.  
It is  easy to appreciate the importance of ANN techniques for setting exclusion limits at the Tevatron. 
For example, in a data sample analyzed passing first  cut selection in Ref.~\cite{DOnote}, 
the data model employed there predicts  only $\sim 12$ signal versus  $\sim 337$  
background $e^{\pm}\mu^{\mp}$  events.

Given the sensitivity of the gluon fusion cross-section to higher order  effects, it  is important to 
establish that the sophisticated methods used in the Tevatron analysis~\cite{Phenomena:2009pt} account  
for these effects within the  estimated  uncertainties.
Already the first cut selection may change the relative importance of higher order corrections for 
signal and background cross-sections with respect to the perturbative  patterns observed 
in the inclusive cross-sections. A complete simulation of  the experimental analysis  with NLO and NNLO perturbative 
corrections  is not practically  feasible.  In this paper,  we aim to  provide precise predictions for the 
production cross section in conditions close to those of the actual experiments by combining  knowledge of higher-order 
effects from fixed-order perturbation theory and parton shower event generators.  

First, we provide fixed-order predictions which are sufficiently detailed to permit
assessment of the sensitivity of the acceptance due to selection cuts.  
For such a  study, fully differential cross-sections of the  signal at  NNLO are required.  
Unlike NLO computations, NNLO differential calculations are a rarity due to their 
substantial technical complications.  The first differential distribution at NNLO was  
computed in 2003~\cite{Anastasiou:2003yy,Anastasiou:2003ds}, and  fully 
differential NNLO cross-sections appeared in  2004~\cite{Anastasiou:2004qd,Anastasiou:2004xq}.  
At an electron-positron collider NNLO differential cross-sections  are known only for 
two~\cite{Anastasiou:2004qd,Weinzierl:2006yt} and three jet production~\cite{Ridder:2009dp,GehrmannDeRidder:2008ug,Dissertori:2007xa,GehrmannDeRidder:2007hr,GehrmannDeRidder:2007bj,Weinzierl:2009nz,Weinzierl:2009ms,Weinzierl:2008iv} cross-sections. 
At hadron colliders  fully differential cross-sections have been computed only 
for Higgs production in gluon fusion~\cite{Anastasiou:2004xq,fehip,Catani:2007vq,Anastasiou:2007mz,hnnlo}, 
and the Drell-Yan process~\cite{Melnikov:2006kv,Melnikov:2006di,Catani:2009sm}. 

In this paper we compute  accepted cross-sections and  kinematic distributions at NNLO using the 
programs  FEHIP~\cite{fehip,Anastasiou:2007mz} and HNNLO~\cite{hnnlo}. We find an excellent  agreement  between the 
predictions of the two programs. We remark that the methods  used~\cite{Anastasiou:2003gr,Catani:2007vq}  for 
constructing these fully differential  NNLO programs are  independent and very different in their conception. 
Both programs  produce kinematic distributions in the form of bin histograms. All NNLO results for cross-sections and 
kinematic distributions calculated with both programs and presented here are in agreement  within the  expectations  
of statistical integration errors.   
The NNLO acceptance of the selection cuts can be directly  compared with the predictions from the modeling  of 
data as it has been performed by the  CDF~\cite{CDFnote} and {D\O\ }~\cite{DOnote} collaborations. 
The kinematic  distributions we present here  are input for their ANN analyses.  

Two selection cuts require special care in estimating their acceptance for the 
Tevatron studies: ({\it i}) a jet veto on two or more central jets and ({\it ii}) isolation of leptons from hadronic activity.
These cuts are used in {D\O\ }~\cite{DOnote} in order to define the entirety of  
the data sample and at CDF~\cite{CDFnote}  that part of the sample  with a potential Higgs boson signal.
At CDF~\cite{CDFnote}, a  further division of the data sample into zero- and one-jet multiplicities is  made. 
The relative magnitude of the perturbative  corrections at NLO and NNLO with respect to LO (\K-factor) 
is smaller after applying  selection cuts.

The same  observation  was made in earlier NNLO studies ~\cite{Anastasiou:2007mz,hnnlo} 
of the process ${\rm pp}\to\higgs\to\WW\to\ell\nu\ell\nu$ at LHC center-of-mass energy of 14 TeV, with 
similar cuts, such as  a jet-veto. In a separate paper~\cite{Anastasiou:2008ik}, these NNLO predictions 
at the LHC were compared to  results obtained from ({\it i}) a resummation of logarithms in 
the transverse momentum of the Higgs boson at NNLL 
accuracy~\cite{Bozzi:2005wk} and ({\it ii}) the Monte-Carlo event generator \mcnlo~\cite{mcnlo}. 
They were found to be in good agreement with each other 
over the phase-space regions singled out by the event
selection cuts. On the contrary, a fixed-order calculation at next-to-leading
order (NLO) accuracy provided a rather poor approximation for the required distributions
and cut efficiencies.

In this paper we compare the selection cut acceptance and  the shape of 
kinematic distributions for leptons at NNLO with the event generators 
\mcnlo~\cite{mcnlo}, \HERWIG~\cite{herwig}, and \PYTHIA8~\cite{pythia8} for 
Tevatron collisions. 

The simulation of the gluon fusion process in Refs.~\cite{CDFnote, DOnote}, is performed with \PYTHIA6~\cite{pythia}.   
In these  analyses, the uncertainty in the acceptance after selection cuts is estimated with other means rather than  
a direct NNLO calculation. In the CDF analysis~\cite{CDFnote},  
\PYTHIA~  events  are  reweighted~\cite{Davatz:2004zg,Davatz:2006ut}  to match either 
the (N)NLO Higgs \pt\ or the NNLO Higgs rapidity spectrum~\cite{fehip,hnnlo}. 
The systematic uncertainty in the acceptance is  computed from the differences between the original \PYTHIA~ and 
the reweighted versions. In D\O, acceptance  uncertainties are  estimated by comparing the spectrum of \PYTHIA~ with 
that of other generators, such as \mcnlo~\cite{mcnlo} and \SHERPA~\cite{sherpa}.  We  believe that neither of the 
two methods can substitute for a direct  comparison with the acceptance at NNLO. We will discuss this point in
Section~\ref{sec:uncertainties}.

Multivariate techniques and distributions of ANN variables have so far been ``terra incognita''  for 
theoretical calculations at higher orders in perturbation theory. To the best of our knowledge, there has  been no 
calculation of NLO and NNLO corrections for such observables. 
So far, the systematic  uncertainty due to higher order effects on the shape of ANN distributions has been  estimated  
indirectly.  ANNs construct  
composite output variables that maximize  the differences between signal and background cross-sections.
An error estimate  on the shape of the ANN output distribution is obtained  by varying the input kinematic distributions  within their uncertainty range.   However, if the input variables have a proper definition at  the parton level,  there is no obstacle  to computing the corrections directly at fixed order in perturbation theory in the same fashion as  for any other simple 
partonic  variable.  This should provide  a more reliable theoretical estimate of the uncertainty on the ANN distribution. 

We demonstrate such a calculation  of an ANN output  distribution through NNLO in this paper.  We train an ANN with a  \PYTHIA~ simulated  data sample that satisfies the selection in Ref.~\cite{Aaltonen:2008ec}, which is a similar  but somewhat simpler 
selection than that in Refs.~\cite{DOnote,CDFnote}.  
As input we use kinematic distributions of the leptons in the final state. We have deliberately refrained from using 
variables such as the transverse  momentum of the Higgs boson or  jets, since these  distributions may differ  
substantially in event generators, and they are not defined in their full physical range  in fixed order perturbation 
theory.   We compare the predictions  of  $\PYTHIA$, $\mcnlo$ and $\HERWIG$ for this ANN output, and 
find reasonable agreement if the observed discrepancies at the cut selection levels are already accounted 
for. 

The organization of the paper is as follows: We first present in Section~\ref{sec:inclusive} our predictions for the
inclusive cross section at various orders of perturbation theory, then in Section~\ref{sec:selection} we define the 
experimental observables to which selection cuts are applied.  Section~\ref{sec:uncertainties} is devoted to a
discussion of the Higgs $\pt$ spectrum and of the jet multiplicities.
In Section~\ref{sec:signal} we
give the results for the accepted cross section, i.e.\ after applying cuts on the various
observables, and discuss the impact of the higher order corrections and scale
variations on the selection efficiency. Next we present detailed comparisons
of kinematical distributions, calculated at different orders of perturbative QCD (Section~\ref{sec:kinematics_nnlo}) and
by using parton shower Monte Carlo models (Section~\ref{sec:comparison}). In Section~\ref{sec:ANN}
we compare for the first time fixed-order and parton shower Monte Carlo predictions for 
the output of an artificial neural network (ANN) similar to that used by the experimental groups.
We comment on the stability and accuracy of these perturbative predictions, depending on the type of  
input variables to the ANN.  Our conclusions are summarized in Section~\ref{sec:TheEnd}.

\section{Inclusive cross section}\label{sec:inclusive}

Recent updates on the inclusive cross section for Higgs boson production at hadron colliders
have been presented in Refs.~\cite{Anastasiou:2008tj,deFlorian:2009hc}.

In Ref.~\cite{Anastasiou:2008tj}, the fixed order NNLO 
cross-section~\cite{Harlander:2002wh,Anastasiou:2002yz,Ravindran:2003um} is recomputed 
with MSTW2008 parton densities~\cite{Martin:2009iq}.  Two-loop electroweak corrections 
from~\cite{Actis:2008ug,Actis:2008ts} and exact finite top and bottom 
mass effects ~\cite{Spira:1995rr,Anastasiou:2006hc} are included. Mixed QCD electroweak effects  are  
taken into account by computing the  relative magnitude of the correction  with respect to the leading 
two-loop electroweak contribution by means of an effective theory. The central value of the cross-section is
determined at a factorization and renormalization scale $\mu = \mh/2$.   The  scale variation 
error for  Higgs mass values $160 - 170\, \GeV$ in this fixed order calculation is 
$+7\%, -11\%$. The corresponding  parton density error  is  $\sim \pm 10\%$. 

In Ref.~\cite{deFlorian:2009hc}, the NNLL calculation~\cite{Catani:2003zt} with the appropriate 
matching to the fixed order NNLO result~\cite{Harlander:2002wh,Anastasiou:2002yz,Ravindran:2003um}
is repeated using the MSTW2008 parton densities~\cite{Martin:2009iq}.  Two-loop electroweak corrections 
from~\cite{Actis:2008ug,Actis:2008ts} and exact finite top and bottom mass effects ~\cite{Spira:1995rr} are 
included. Mixed QCD electroweak effects  are  taken into account by multiplying the inclusive two-loop electroweak 
contribution~\cite{Actis:2008ug,Actis:2008ts} with the QCD \K-factor.
The central value of the cross-section is determined at a factorization and renormalization scale 
$\mu = \mh$.   The  scale variation error for  Higgs mass values $160 - 170\, \GeV$ in this resummed
calculation is $+9\%, -8\%$. The corresponding  parton density error is $\sim \pm 8\%$.

The central values of the cross-sections in Refs~\cite{Anastasiou:2008tj,deFlorian:2009hc}  agree within
a percent, and they have been used in setting the exclusion limits in~\cite{Phenomena:2009pt}.  
CDF assigns  a  global theoretical uncertainty  on the inclusive  cross-section of $\pm 12\%$~\cite{CDFnote}, 
adding in quadrature  a $\pm 11\%$ scale variation error and a $\pm 5\%$ parton density error. 
In our opinion, the combination of the two errors  in  quadrature  requires  further  justification.
{D\O\ } assigns  a  somewhat smaller theoretical uncertainty of $\pm 10\%$ to the inclusive Higgs  boson 
cross-sections.
Even justifying a combination in quadrature of the uncertainties from parton densities and scale variations,
the uncertainties on the total rate appear underestimated,
compared to those of Refs.~\cite{Anastasiou:2008tj,deFlorian:2009hc}.

The main goal of the present paper is to assess  the theoretical uncertainties  on the 
shapes of kinematic distributions and  the acceptance  after selection cuts. 
We have used  MRST2001 PDFs at LO and MRST2004 PDFs at NLO and NNLO.
All the fixed-order results in this paper have been obtained  independently with the 
FEHiP~\cite{Anastasiou:2004xq,fehip,Anastasiou:2007mz,Anastasiou:2008ik} and 
HNNLO~\cite{Catani:2007vq,hnnlo} programs, by first calculating  \K-factors in the limit of a very 
heavy top-quark and then by multiplying these \K-factors with the exact leading order gluon 
fusion cross section for Higgs production via a top-quark (ignoring bottom contributions). The total 
width is  computed using  the program HDECAY~\cite{hdecay}.  ${\cal O}(\alpha)$ electroweak corrections for the 
Higgs partial decay to leptons~\cite{Bredenstein:2006rh} have also been ignored.  This is justified for 
studies  of shapes and acceptances.  
\begin{table}[h]
  \begin{center}    
    \begin{tabular}{|l||c|c|c||c|c|}
      \hline
      $\sigma_\mathrm{inc}\;\;[\fb]$ & LO & NLO & NNLO & $K^\mathrm{NLO}$ & $K^\mathrm{NNLO}$\\\hline\hline
      $\mu=\mh/2$ & $1.998\pm0.003$ & $4.288\pm0.004$ & $5.252\pm0.016$ & $2.149\pm0.008$ & $2.629\pm0.009$ \\
      $\mu=\mh$ & $1.398\pm0.001$ & $3.366\pm0.003$ & $4.630\pm0.010$ & $2.412\pm0.002$ & $3.312\pm0.008$\\
      $\mu=2\,\mh$ & $1.004\pm0.001$ & $2.661\pm0.002$ & $4.012\pm0.007$ & $2.651\pm0.008$ & $3.996\pm0.008$\\
      \hline
    \end{tabular}
  \end{center}
  \caption{Inclusive cross sections for $\mh=160\,\GeV$, at various orders in perturbation theory and for different
  scale choices. The \K-factors are defined in the text; LO=Leading order.
    \label{tab:inccross}}
\end{table}

The cross-section for $\ppbar\to\higgs\to\WW\to\ell\nu\ell\nu$, 
with no cuts applied and using the setup described in the last  paragraph, 
is presented in Table~\ref{tab:inccross} for illustration.  We  have chosen to study 
the signal cross-section for a Higgs mass value $\mh = 160\, \GeV$, and  we vary the  
renormalization ($\mur$) and factorization ($\muf$) scales simultaneously in the 
interval  $\mu=\muf=\mur\in[\mh/2,2\,\mh]$. In the same  Table, we  also present 
the \K-factors for the inclusive cross section, 
\begin{equation}
K^{\mathrm{(N)NLO}}(\mu)=\frac{\sigma^{\mathrm{(N)NLO}}(\mu)}{\sigma^{\mathrm{LO}}(\mu)}.
\end{equation}
The perturbative  corrections  are very large, and there is still a substantial $23 - 50\%$ increase, 
depending on the  scale choice, of the cross-section at NLO  upon including the NNLO corrections.  
The  smallest \K-factors occur  for smaller values  of the renormalization and  factorization scale.

\section{Observables and selection cuts}
\label{sec:selection}

The very sophisticated  and  complex experimental analyses of the two Tevatron collaborations cannot be 
reproduced fully at a perturbative level.   In addition, there are important differences between the CDF 
and {D\O\ }  analyses, both at the event selection level and in the usage of ANNs, 
and they evolve with time and the acquisition of more data.  Nevertheless, we believe the
analysis presented here captures the essential features that allow us to study the likely
sensitivity of the results to higher order effects.

Our particle and event selection follows along the lines of the CDF analysis of Ref.~\cite{Aaltonen:2008ec} and proceeds  through the following steps:
\begin{enumerate}
\item
Lepton selection: in the CDF experiment, the experimental acceptances for electrons and
muons are different. In this publication, we only consider the final state with two muons,
simulating only the muon acceptance. The differences  in the CDF cuts for muons and
electrons are rather geometric, and should not alter  the convergence
pattern of the  perturbative corrections.   In our analysis,
one of the final-state leptons (the `trigger lepton')
must have a transverse momentum $\pt>20\,\GeV$ and pseudo-rapidity $|\eta|<0.8$.
In order to  pass a further lepton selection,  a second lepton must be found with
$\pt>10\,\GeV$ and $|\eta|<1.1$. 
\begin{enumerate}
\item Two opposite-sign leptons have to be found, fulfilling the requirements discussed above.
\item Both leptons have to be isolated, i.e. the additional transverse energy in a cone with radius $R=0.4$ 
         around the lepton has to be smaller than 10\,\% of the lepton transverse momentum.
\item In order to reduce the background from \bq~resonances, 
the invariant mass of the lepton pair has to be $\mll>16\,\GeV$.
\end{enumerate}
\item   We define the missing transverse energy (MET) as the vectorial sum of the transverse momenta of the two neutrinos.
We define the variable $\mathrm{MET}^*$ as 
\begin{equation}
\mathrm{MET}^*=\left\{\begin{array}{l} \mathrm{MET}\;\;\;\;\;\;\;\;\;\;\;\;,\;\phi\geq\pi/2  \\ \mathrm{MET}\times\sin{\phi}\;,\;\phi<\pi/2 \end{array}\right.,
\end{equation}
where $\phi$ is the angle in the transverse plane between MET and the nearest charged lepton or jet. 
We require $\mathrm{MET}^*>25$ GeV, which suppresses the background from Drell-Yan lepton pairs 
and removes contributions from mismeasured leptons or jets.

\item In order to suppress the $t{\bar t}$ background, we apply a veto on the number of jets in the event. 
Jets are found  using the  \kt-algorithm~\cite{Catani:1993hr,Ellis:1993tq} with parameter $R=0.4$.  A jet must have
$\pt >15\,\GeV$ and $|\eta|<3.0$.  Events are only accepted if there is no more than one such jet.

\end{enumerate}

The jet veto that we apply here is different from that used in our corresponding LHC 
studies~\cite{Anastasiou:2007mz,Anastasiou:2008ik,hnnlo}, where all 
events with any number of central jets  with a \pt\ higher than  a  certain minimum value 
are  vetoed. The cuts in the present study allow  for events with a single high-\pt\  jet. This type of jet veto is 
used in the {D\O\ }  analysis~\cite{DOnote} in order  to define the data sample  with a potential 
Higgs signal. A  stricter jet veto  is applied in the CDF analysis~\cite{CDFnote}, where three data 
samples are defined according to whether events  have zero, one, or more central jets.  

\section{Higgs $\pt$ spectrum and jet multiplicities}
\label{sec:uncertainties}

\begin{figure}[th]
  \begin{center}
    \includegraphics[width=0.48\textwidth]{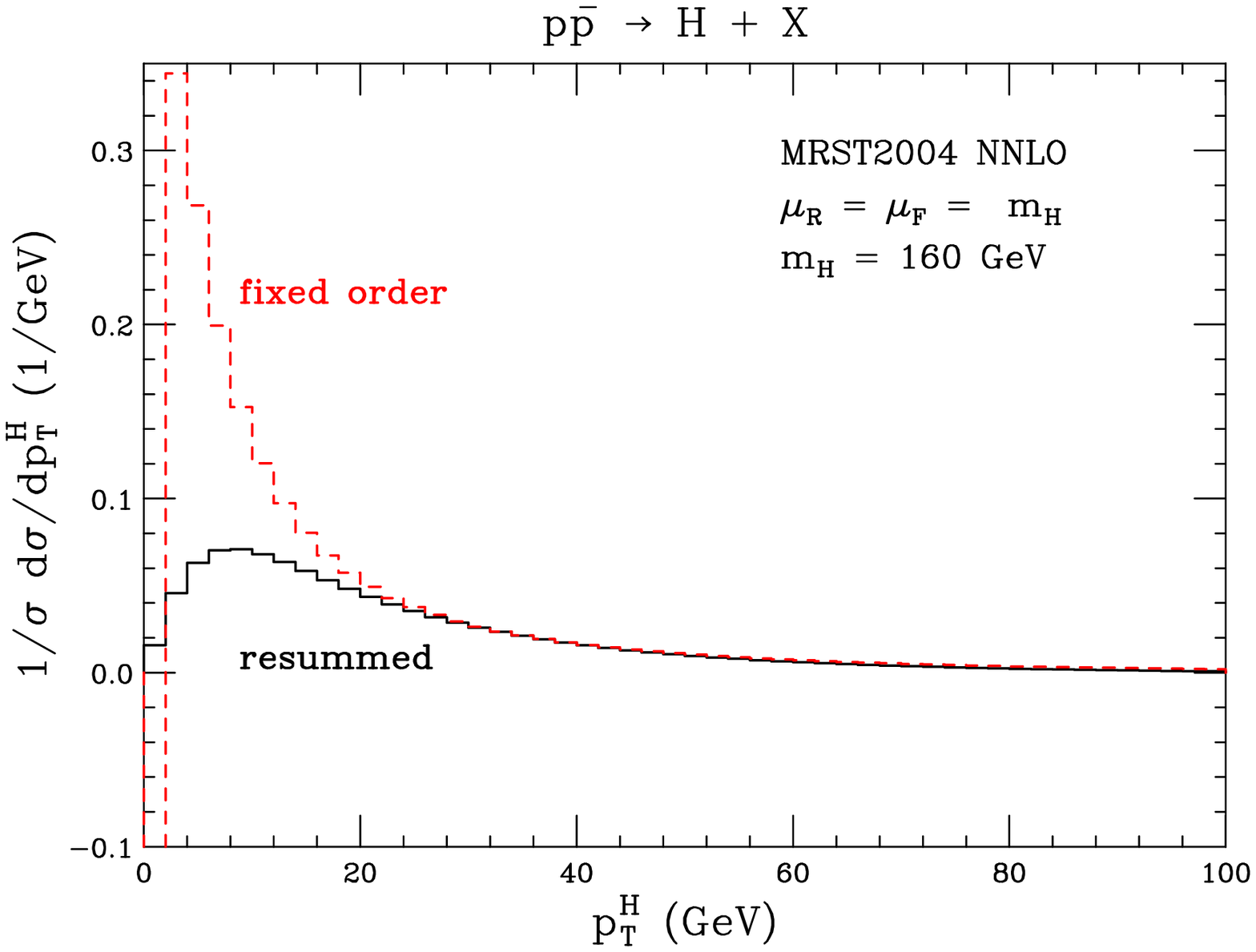}
    \includegraphics[width=0.48\textwidth]{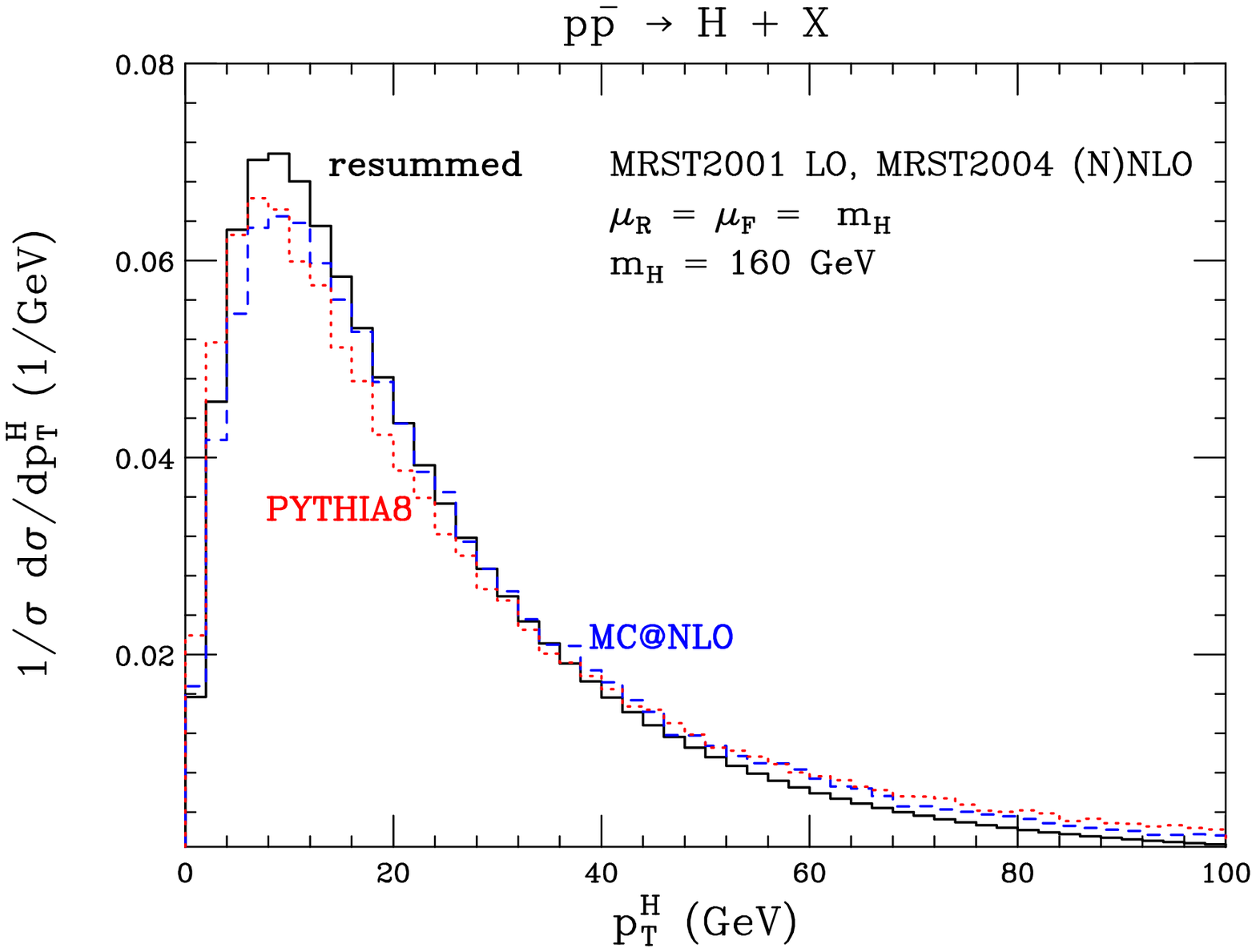}
  \end{center}
  \caption{\label{fig:pthiggs}
    On the left figure, we show the normalized transverse momentum distributions 
    for $\mh=160\,\GeV$ and  $\mu=\mur=\muf = \mh$ using NNLO fixed-order perturbation
    theory and  the  resummed calculation of Ref.~\cite{Bozzi:2005wk}.  
    On the right figure, the same distribution is shown for \mcnlo, \PYTHIA8, and the 
    calculation of Ref.~\cite{Bozzi:2005wk}.}
\end{figure} 

One of the most important distributions for a Higgs boson produced at hadron colliders
is its transverse momentum spectrum.
A good description of the $\pt$ spectrum implies a good
understanding of the QCD radiation recoiling against the Higgs.

It is well known that the Higgs $\pt$ spectrum is not physical when computed at fixed order,
since it diverges to $+\infty$ or $-\infty$ at any fixed order in $\alpha_s$.
When $\pt\ll \mh$ large logarithmic contributions
of the form $\alpha_s^n\,\ln^m \mh/\pt$ appear that must be resummed to all orders.
In Ref.~\cite{Bozzi:2005wk} the resummation of these logarithmically enhanced terms has been performed analytically up to NNLL accuracy,
and the result has then been matched
to the fixed order calculation up to ${\cal O}(\alpha_s^4)$.
The integral of the ensuing spectrum coincides with the total NNLO cross section. 
In Fig.~\ref{fig:pthiggs} (left)
we compare the normalized $\pt$ spectrum of the Higgs computed at fixed order,
to the one obtained with the numerical program of Ref.~\cite{Bozzi:2005wk}.
We see that the fixed order calculation diverges to $-\infty$ as $\pt\to 0$.
The resummed calculation is instead well behaved as $\pt\to 0$. We also note that the two calculations are
in good agreement for $\pt$ larger than about 30 GeV.

Standard Monte Carlo event generators effectively
perform the transverse-momentum resummation, and thus obtain a well behaved $\pt$ spectrum.
A comparison of the $\pt$ spectra from {\PYTHIA~} and  {\mcnlo~} to the resummed calculation
 is presented in Fig.~\ref{fig:pthiggs} (right). We notice that the {\PYTHIA~} spectrum is softer
than those of {\mcnlo~} and the NNLL calculation.
This difference is more marked at LHC energies~\cite{Balazs:2004rd}. 

We now comment on the procedure used by CDF and D\O~to estimate the systematic uncertainties on the signal acceptance.
The simulation of the gluon fusion process in Refs~\cite{CDFnote, DOnote}, is performed with \PYTHIA~\cite{pythia}.   
In these  analyses, the uncertainty in the acceptance after selection cuts is estimated with different methods.
In the CDF analysis~\cite{CDFnote},  
\PYTHIA~  events  are  reweighted~\cite{Davatz:2004zg,Davatz:2006ut}  to match either 
the (N)NLO Higgs \pt\ or the NNLO Higgs rapidity spectrum~\cite{fehip,hnnlo}. 
The systematic uncertainty in the acceptance is  computed from the differences between the original \PYTHIA~ and 
the reweighted versions. In D\O, acceptance  uncertainties are  estimated by comparing the spectrum of \PYTHIA~ with 
that of other generators, such as \mcnlo~\cite{mcnlo} and \SHERPA~\cite{sherpa}.  We  believe that neither of the 
two methods can substitute for a direct comparison with the acceptance at NNLO.
 
The reweighting technique used by CDF is based on the NNLO $p_T$ spectrum of the Higgs boson.
As shown in Fig.~\ref{fig:pthiggs}, the fixed-order Higgs $\pt$ spectrum is divergent as $\pt\to 0$.
A reweighting procedure based on the NNLO Higgs $\pt$
spectrum makes sense only if the signal rate is integrated over a reasonably large
$\pt$ region (as is done, for example, in Ref.~\cite{Davatz:2006ut}). In this respect, 
it would be better to reweight using a resummed calculation \cite{Bozzi:2005wk}, 
or to use different kinematic variables.

Note also that a reweighting based on the Higgs $\pt$ and rapidity is insensitive to the jet multiplicity of the event,
which is  used in order to divide the data sample. Another important aspect is that
the relative weight of the one- and two-jet sample is enhanced by the CDF cuts, with respect to the zero-jet sample.
If a jet is  required in all events, the ${\cal O}(\alpha_s^4)$ calculation includes  
matrix elements  through NLO only.  If two jets are required, only LO matrix-elements are taken into account.
An NLO  prediction for 1-jet samples and  a LO prediction for 2-jet samples~\footnote{
An NLO calculation for  $H+2$ jets via  gluon fusion can be found in~\cite{Campbell:2006xx}.}  
can be obtained with FEHIP and HNNLO using the corresponding  NLO and LO $\alpha_s$ and parton densities.  
More importantly,
we find it inconsistent to use the theoretical uncertainty from the inclusive NNLO gluon fusion 
cross-section  as  the uncertainty of the samples with defined jet multiplicities other than zero. 

To demonstrate this point further, we follow  an analogous  procedure as in 
\cite{CDFnote} and divide  the signal cross-section into three bins  according to the 
number of  central jets. Jets are  defined using the $\kt$-algorithm, with minimum 
$\pt = 15$ GeV and  maximum rapidity $| \eta | < 2$.  We compute the inclusive cross-sections 
for $\mh = 160$ GeV and vary the renormalization and factorization scale
simultaneously in the interval $[\mh/2, 2\mh]$. In Table~\ref{tab:jetbins} 
we compute the three bin cross-sections, using either NNLO or NLO or LO parton densities and $\alpha_s$ evolution from the recent MSTW2008 fit \cite{Martin:2009iq}.
\begin{table}[h]
  \begin{center}    
    \begin{tabular}{|l||c|c|c||}
      \hline
      $\sigma\;[\fb]$ & LO (pdfs, $\alpha_s$) & NLO (pdfs, $\alpha_s$) & NNLO (pdfs, $\alpha_s$)\\\hline\hline
        0-jets  &$  3.452^{+7\%}_{-10\%}  $&$ 2.883^{+4\%}_{-9\%}  $&$ 2.707^{+5\%}_{-9\%} $\\ \hline 
        1-jet   &$  1.752^{+30\%}_{-26\%}  $&$ 1.280^{+24\%}_{-23\%}  $&$ 1.165^{+24\%}_{-22\%} $\\ \hline
 $\geq$ 2-jets  &$  0.336^{+91\%}_{-44\%}  $&$ 0.221^{+81\%}_{-42\%}  $&$  0.196^{+78\%}_{-41\%}  $\\
      \hline \hline
    \end{tabular}
  \end{center}
  \caption{
\label{tab:jetbins}
Inclusive cross sections in the different jet bins.}
\end{table}

The total cross-section  with NNLO pdfs varies  around the  default scale value $\mu=\mh$ by $\pm 14\%$.
From Table~\ref{tab:jetbins} we see that about $66.5\%$ of the events contain zero jets, 
$28.6\%$ one jet only, and $4.9\%$ contain more than one jets.  
 Notice, however,  that the scale variation in the three jet bins is significantly different and deteriorates with increasing jet multiplicity.  This is a  consequence of the fact that in the 1-jet and 2-jet bins the fixed order calculation is only accurate through NLO and LO, respectively.  The resulting scale dependence of the inclusive cross section is made up as follows:
\begin{equation}
\label{eq:jetbinvar_inc}
\frac{\Delta N_{{\rm inc}}({\rm scale})}{N_{{\rm inc}}} =  
66.5\% \cdot \left({^{+5\%}_{-9\%}} \right) 
+28.6\% \cdot \left({^{+24\%}_{-22\%}} \right) 
+4.9\% \cdot \left({^{+78\%}_{-41\%}} \right) = \left({^{+14.0\%}_{-14.3\%}} \right)  
\end{equation}

The application of different selection cuts  in the three  jet bins  leads to a theoretical error estimate of the number of  signal events which is different from the theoretical error of the inclusive NNLO cross-section. Specifically, from Tables 1-3  of Ref.~\cite{CDFnote} we observe that, after preselection,  $60\%$ of gluon fusion events belong to the 0-jets bin, $29\%$ to the 1-jet bin, and $11\%$ to the 2-jet bin. 

We now examine how this modification of  the jet multiplicities  with the experimental cuts affects the scale variation for the total number of 
events. With the exception of the jet-veto, all other cuts  used in the CDF  preselection~\cite{CDFnote} do not  affect the  scale variation of the total cross-section significantly. We  can then estimate the scale variation of the total number of signal 
events using the scale-variations  for each jet-multiplicity in Table~\ref{tab:jetbins} 
and  the  expected composition of jet-multiplicities for the signal~\cite{CDFnote}. 
Using NNLO pdf's and NNLO $\alpha_s$ evolution for all jet bins, we find that: 
\begin{equation}
\label{eq:jetbinvar_nnlo}
\frac{\Delta N_{{\rm signal}}({\rm scale})}{N_{{\rm signal}}} =  
60\% \cdot \left({^{+5\%}_{-9\%}} \right) 
+29\% \cdot \left({^{+24\%}_{-22\%}} \right) 
+11\% \cdot \left({^{+78\%}_{-41\%}} \right) = \left({^{+18.5\%}_{-16.3\%}} \right)  
\end{equation}
The  resulting scale variation is therefore larger than the corresponding 
scale  variation of $\pm 14\%$ for the inclusive  cross-section. 

Notice that in Eq.~\ref{eq:jetbinvar_nnlo} we used a scale variation for the  one-jet and two-jet 
bins corresponding to NNLO pdfs and $\alpha_s$ evolution. A more consistent approach would be to estimate the number of events in the 1-jet and 2-jet bins using NLO and LO pdfs and 
$\alpha_s$ evolution correspondingly.  In this way we  obtain: 
\begin{equation}
\label{eq:jetbinvar_consistent}
\frac{\Delta N_{{\rm signal}}({\rm scale})}{N_{{\rm signal}}} =  
60\% \cdot \left({^{+5\%}_{-9\%}} \right) 
+29\% \cdot \left({^{+24\%}_{-23\%}} \right) 
+11\% \cdot \left({^{+91\%}_{-44\%}} \right) = \left({^{+20.0\%}_{-16.9\%}} \right)  
\end{equation}

The relative population of the jet bins is  very important for the determination of the theoretical error on the total number  of events.  The  contribution of the different jet multiplicities to the total error can be altered  also after the  preselection cuts, since, in general, the independent multivariate methods for discriminating signal from background events in various jet bins should have different discriminating efficiency.  In conclusion, the theoretical 
error  for the number of events  at  various jet multiplicities  should  not be estimated  collectively from the scale variation of the total cross-section.

\section{Signal cross section and  preselection efficiency at fixed order} \label{sec:signal}

We  present  in Table~\ref{tab:accxsec} the LO, NLO, and NNLO cross sections, as obtained with FEHIP and HNNLO 
after applying the selection cuts in Section~\ref{sec:selection}, for a default Higgs boson mass value $\mh = 160 \, \GeV$. 
\begin{table}[h]
  \begin{center}    
    \begin{tabular}{|l||c|c|c||c|c|}
      \hline
      $\sigma_\mathrm{acc}\;\;[\fb]$ & LO & NLO & NNLO & $K^\mathrm{NLO}$ & $K^\mathrm{NNLO}$\\\hline\hline
      $\mu=\mh/2$ & $0.750\pm0.001$ & $1.410\pm0.003$ & $1.459\pm0.003$ & $1.880\pm0.005$ & $1.915\pm0.025$\\
      $\mu=\mh$ & $0.525\pm0.001$ & $1.129\pm0.003$ & $1.383\pm0.004$ & $2.150\pm0.007$ & $2.594\pm0.052$\\
      $\mu=2\,\mh$ & $0.379\pm0.001$ & $0.903\pm0.002$ & $1.242\pm0.001$& $2.383\pm0.008$ & $3.261\pm0.048$\\
      \hline
    \end{tabular}
  \end{center}
  \caption{
\label{tab:accxsec}
Accepted cross sections and \K-factors after the application of all the 
selection cuts for $\mh=160\,\GeV$.}
\end{table}

Comparison of the results of Table \ref{tab:accxsec}  with those of Table \ref{tab:inccross} shows that the impact of 
QCD radiative corrections is significantly reduced  when selection cuts are applied. Indeed, 
for $\mu_F=\mu_R=m_H$ the NLO and NNLO \K-factors are reduced by $11\%$ and $22\%$, respectively. 
As a consequence, the acceptance is also reduced, since it is defined as the ratio
of the cross-section after cuts to the inclusive cross section.
At LO about $38\%$ of the  events are accepted.  At NLO, the efficiency drops to $33\%-34\%$ and at  NNLO to 
$28\% - 31\%$, depending on the  scale choice. 

An important observation is that the scale dependence of the efficiency becomes stronger when we increase  
the  perturbative order. The lepton isolation and jet-veto cuts do not change  the 
LO cross section, since the isolation requirement gives a non-vanishing contribution only at NLO
and the veto on the number
of central jets is effective only beyond NLO.  
We will comment  further on the jet-veto later when comparing to the Monte-Carlo event generators, \PYTHIA, \HERWIG\ and \mcnlo. 

\begin{figure}[h]
  \begin{center}
    \includegraphics[width=0.78\textwidth]{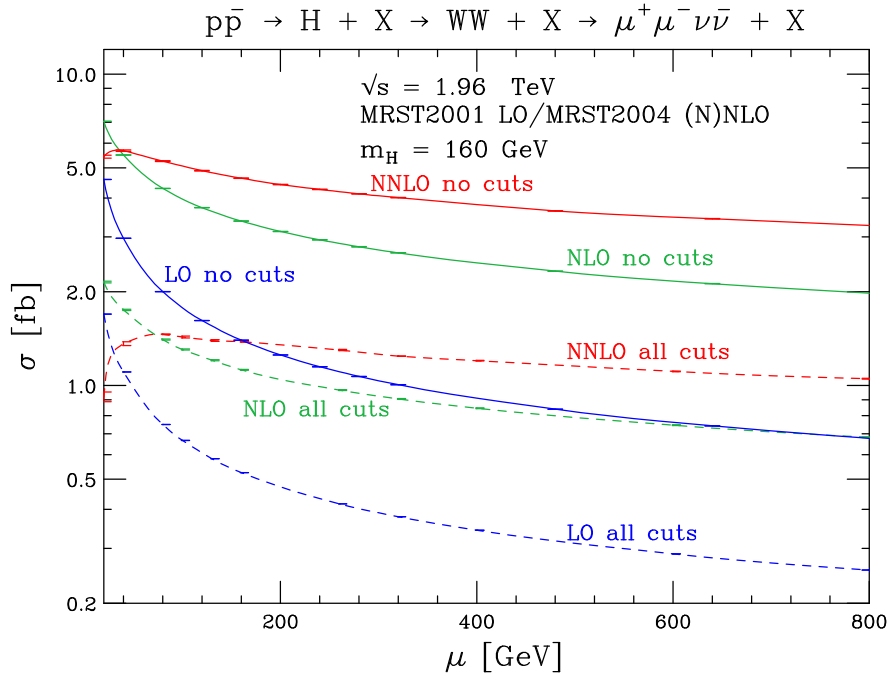}
  \end{center}
  \caption{\label{fig:scalevar}
    Cross sections for $\mh=160\,\GeV$ as a function of the scales
    $\mu=\mur=\muf$ for different orders in fixed-order perturbation
    theory. The solid lines represent inclusive cross sections, the dashed lines cross sections after the application of the selection cuts.}
\end{figure}

The increased scale variation of the acceptance at NNLO is  a reflection of the reduction of the scale variation for the cross section after cuts  are applied.  In Fig.~\ref{fig:scalevar}, we present the cross section as 
a function  of  the renormalization and  factorization scale $\mu = \mu_R = \mu_F$, before and after cuts, 
at LO, NLO, and NNLO.  Before  cuts, the NNLO cross section varies by $27\%$ over the interval 
$\left[ \mh/2, 2\mh\right]$.  This drops to a  $16 \%$ scale  variation after  cuts,
and consequently the ratio varies by $\sim 11\%$.
We note that the scale variation in a  different but also commonly used range 
$\left[ \mh/4, \mh \right]$ is $20\%$ for the inclusive cross-section and $6\%$ for the accepted  cross-section after cuts. In this second  scale  variation range, the accepted cross-section is  
not monotonic and  develops a maximum.

\section{Kinematic distributions at fixed order perturbation theory}
\label{sec:kinematics_nnlo}
We now turn to a more detailed study of the kinematical properties of the accepted events.
The applied cuts provide a rough discrimination of the Higgs signal over the background.
The experimental analysis proceeds further by exploiting the differences in the kinematical distributions between
signal and background and is typically based on Artificial Neural Network (ANN) techniques. 
It is thus important to check that these distributions are stable against radiative corrections.

\begin{figure}[ht]
  \begin{center}
    \includegraphics[width=0.48\textwidth]{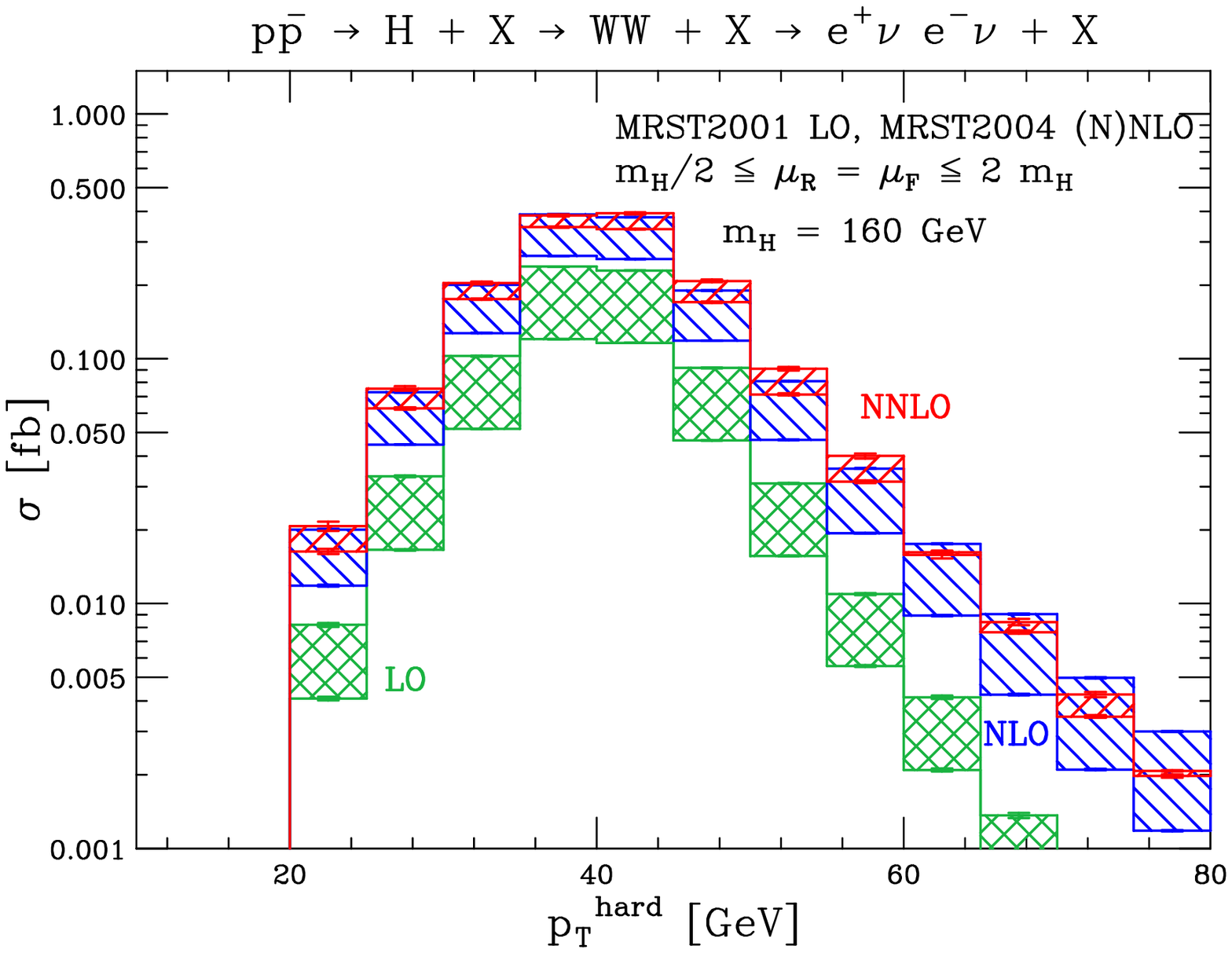}
    \includegraphics[width=0.48\textwidth]{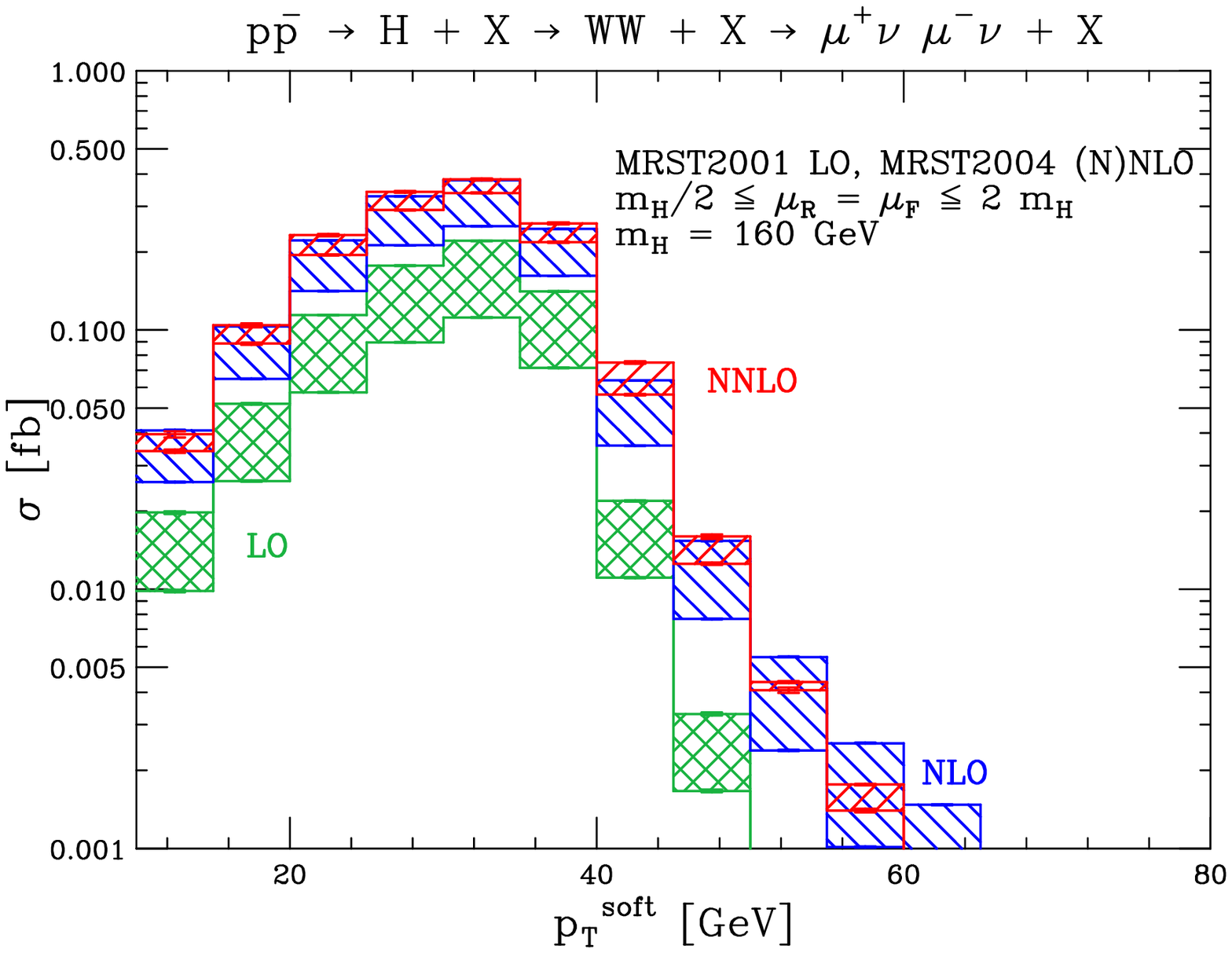}\\
    \includegraphics[width=0.48\textwidth]{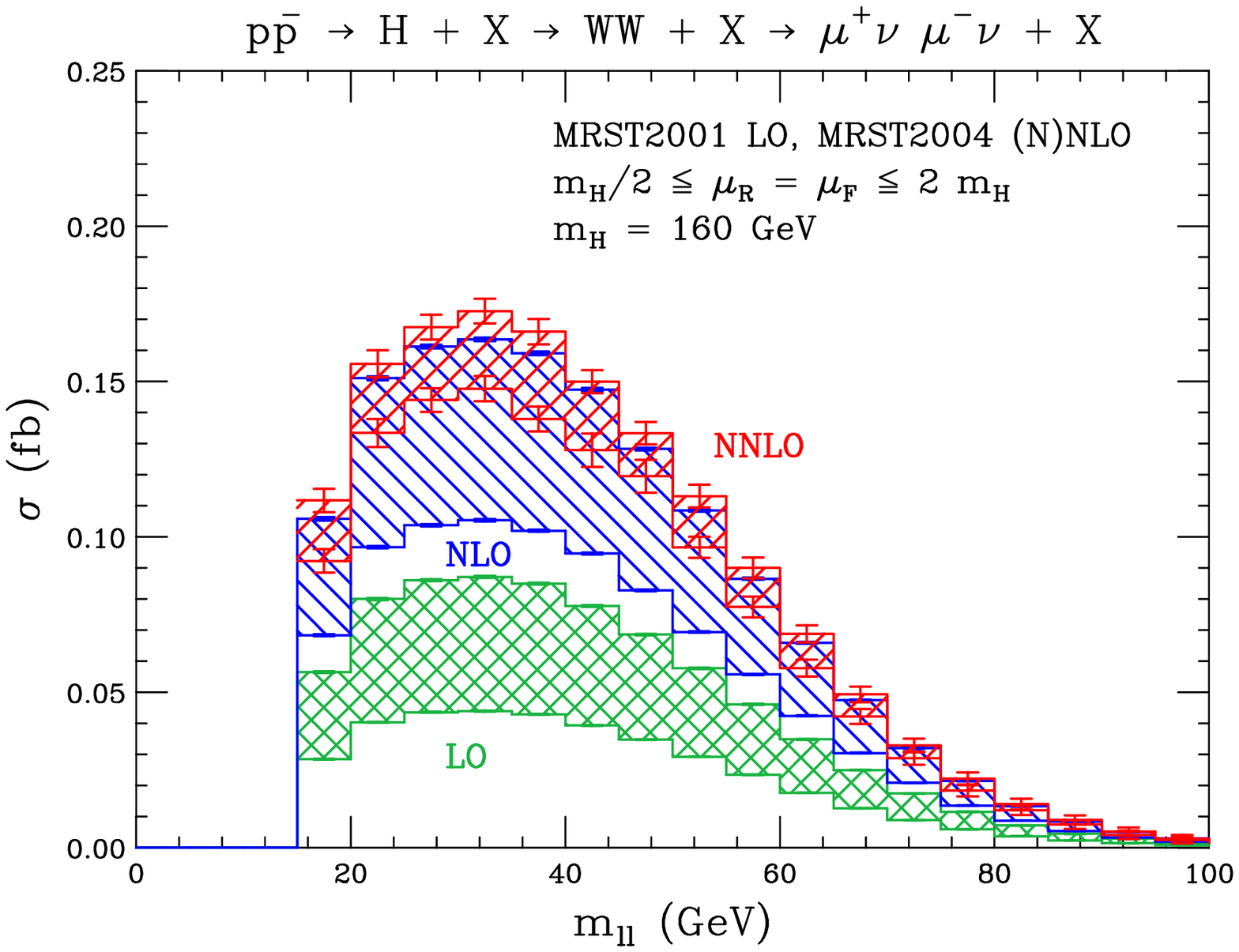}
    \includegraphics[width=0.48\textwidth]{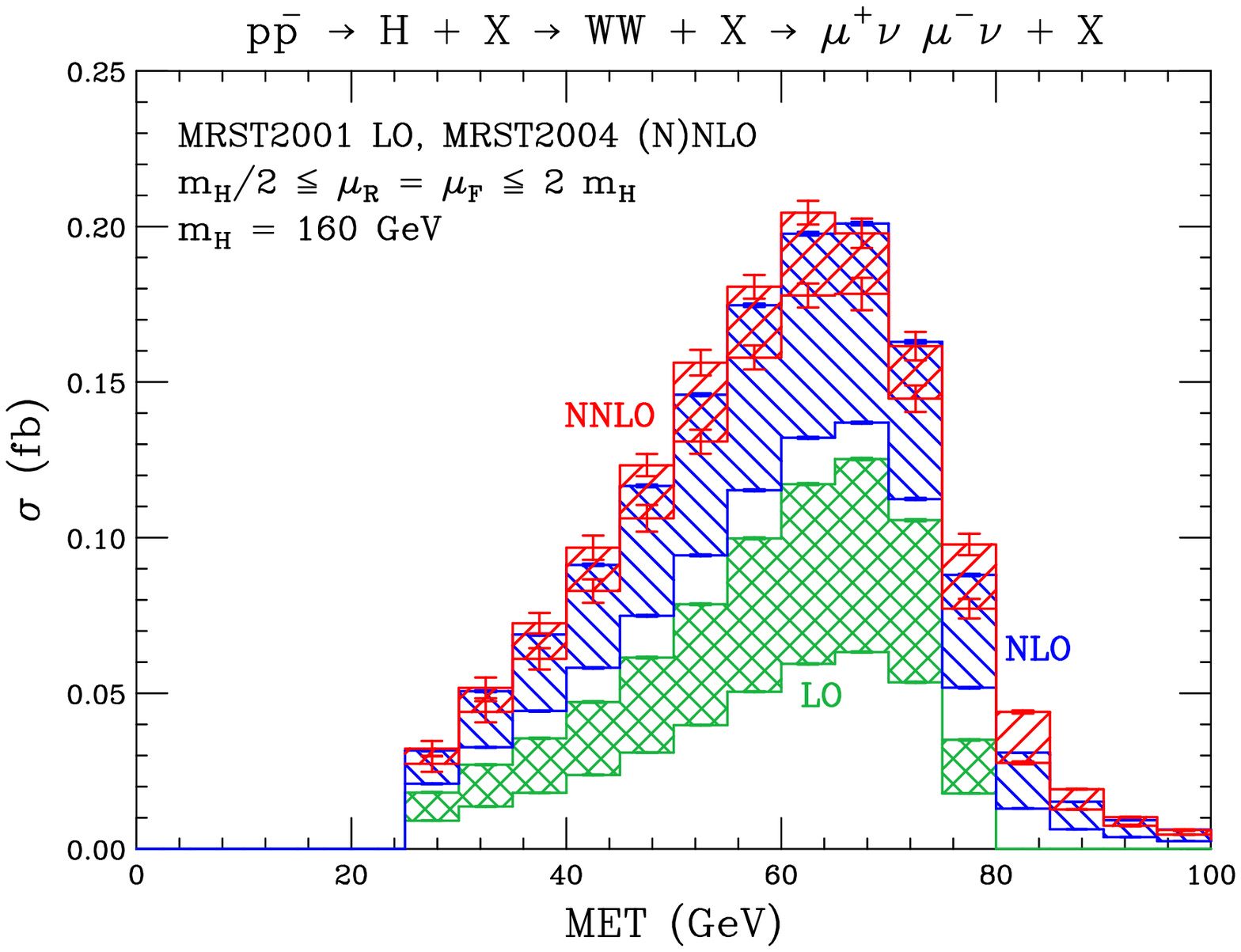}\\
    \includegraphics[width=0.48\textwidth]{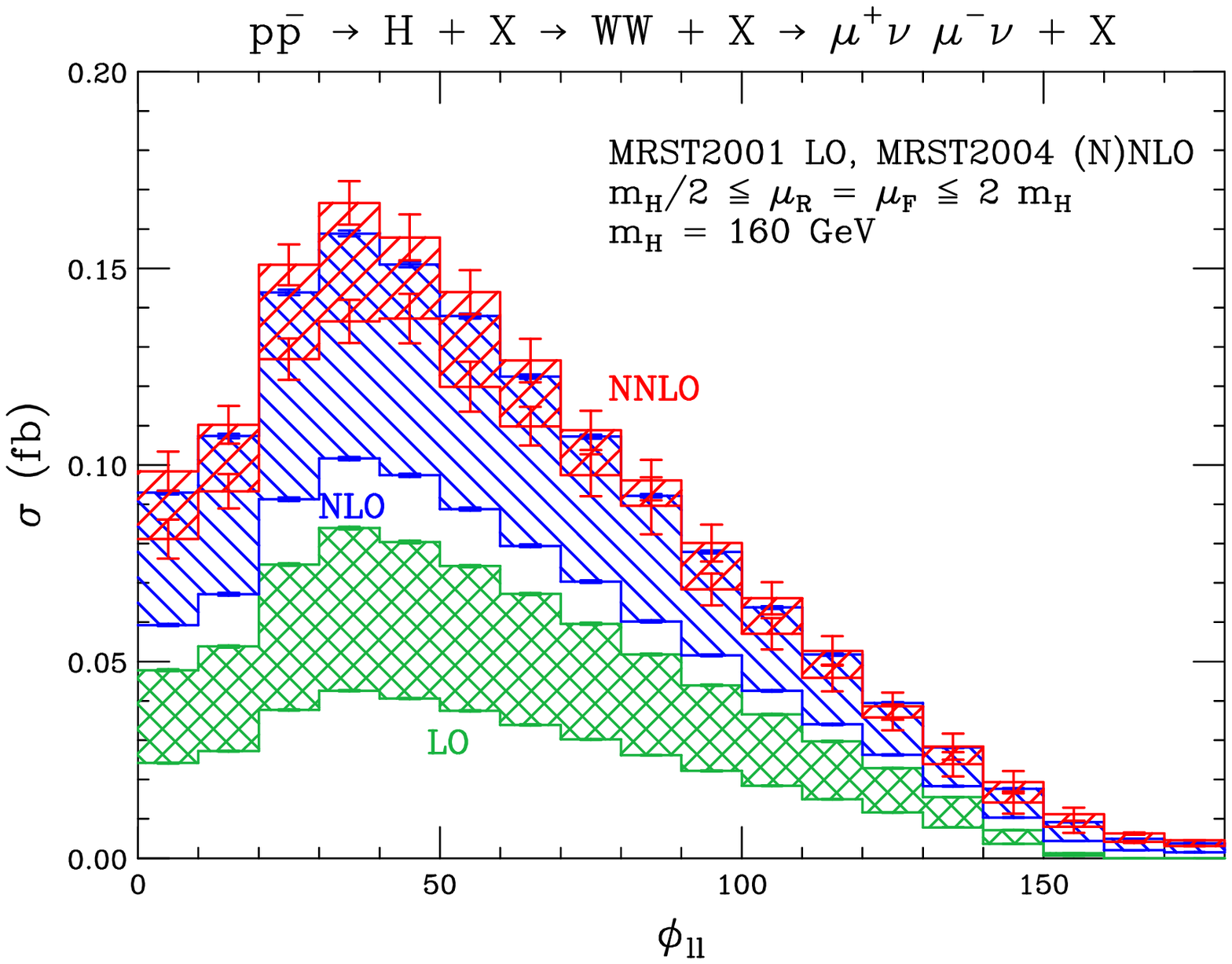}
  \end{center}
  \caption{\label{fig:fixord_distr}
  Kinematic distributions obtained at LO, NLO and NNLO in perturbative QCD. Shown are the transverse
    momentum of the leading ($\pthard$) and trailing lepton ($\ptsoft$), the invariant mass of the two leptons ($m_{ll}$),
    the transverse missing energy (MET) and the azimuthal separation of the two charged leptons in the transverse plane ($\phill$).
}
\end{figure}

In Fig.~\ref{fig:fixord_distr} we present a set of distributions that are commonly used in the 
experimental analysis,  computed at LO, NLO and NNLO. The uncertainty bands are obtained
by varying $\mu_F=\mu_R$ between $m_H/2$ and $2m_H$.
Figures \ref{fig:fixord_distr}(a) and (b) show the transverse momentum spectrum of the leading and
 trailing lepton, $\pthard$ and $\ptsoft$. Figure ~\ref{fig:fixord_distr}(c) shows the invariant
mass distribution of the lepton pair, $m_{ll}$, Fig.~\ref{fig:fixord_distr}(d) shows the MET distribution
and, finally, Fig.~\ref{fig:fixord_distr}(e) shows the azimuthal separation of the two charged leptons in the transverse plane, $\phill$.
Overall, these plots show that the distributions are quite stable when going from NLO to NNLO.
The NNLO band generally lies on top of the NLO band and nicely overlaps with the latter.
The scale uncertainty of the distributions is consistent with that quoted in 
Table~\ref{tab:accxsec} and is about $\pm(6-8)\%$ in the peak region of the distributions.

We  also observe that the perturbative corrections from NLO to NNLO are 
smaller when the scale $\mu = \mh/2$ is used.

\section{Comparison with parton shower event generators}\label{sec:comparison}

In the next step we compare the accepted cross-section and kinematic 
distributions after selection cuts,  obtained in fixed order perturbation theory, with the predictions  
of the parton shower event generators  \HERWIG, \PYTHIA\ and  \mcnlo.  
We consider this as a very important validation for the current  studies at the Tevatron, which 
rely on event generator predictions.  
\begin{table}[h]
\begin{center}
\begin{tabular}{|l|cc|cc|cc|}
\hline
  $\accsigma$ [\fb]          & \multicolumn{2}{c|}{$\mu=\mh/2$} & \multicolumn{2}{c|}{$\mu=\mh$}& \multicolumn{2}{c|}{$\mu=2\;\mh$}      \\
\hline\hline
  LO                         & \multicolumn{2}{c|}
{$0.750\,\pm\,0.001$}  & 
\multicolumn{2}{c|}{$0.525\,\pm\,0.001$}     & \multicolumn{2}{c|}{$0.379\,\pm\, 0.001$}    \\
 $R^{\mathrm{LO}}$(\HERWIG)                      & 
\multicolumn{2}{c|}{$0.571\, \pm \, 0.001$} 
& \multicolumn{2}{c|}{$0.399\, \pm\, 0.001$}      
& \multicolumn{2}{c|}{$0.287\, \pm \,  0.001$}    \\
$R^{\mathrm{LO}}$(\PYTHIA)                      & \multicolumn{2}{c|}{$0.483\,\pm\,0.001$} & \multicolumn{2}{c|}{$0.338\,\pm\,0.001$}      & \multicolumn{2}{c|}{$0.245\,\pm\,0.001$}    \\
\hline\hline
  NLO                        & \multicolumn{2}{c|}{$1.410\,\pm\,0.003$}  & \multicolumn{2}{c|}{$1.129\,\pm\,0.003$}     & \multicolumn{2}{c|}{$0.903\,\pm\,0.002$}    \\
  \mcnlo                     & \multicolumn{2}{c|}{$1.143\,\pm\,0.002$}  & \multicolumn{2}{c|}{$0.906\,\pm\,0.003$}     & \multicolumn{2}{c|}{$0.718\,\pm\,0.002$}    \\
  $R^{\mathrm{NLO}}$(\HERWIG)  & 
\multicolumn{2}{c|}{$1.225\, \pm \, 0.001$} 
& \multicolumn{2}{c|}{$0.962\, \pm\, 0.001$}      
& \multicolumn{2}{c|}{$0.760\, \pm \,  0.001$}    \\
  $R^{\mathrm{NLO}}$(\PYTHIA)  & \multicolumn{2}{c|}{$1.037\,\pm\,0.002$} & \multicolumn{2}{c|}{$0.814\,\pm\,0.002$}      & \multicolumn{2}{c|}{$0.645\,\pm\,0.001$}    \\
  \hline\hline
  NNLO                       & \multicolumn{2}{c|}{$1.459\,\pm\,0.003$} & \multicolumn{2}{c|}{$1.383\,\pm\,0.004$}      & \multicolumn{2}{c|}{$1.242\,\pm\,0.001$}    \\
  $R^{\mathrm{NNLO}}$(\mcnlo)& \multicolumn{2}{c|}{$1.399\,\pm\,0.003$} & \multicolumn{2}{c|}{$1.246\,\pm\,0.003$}      & \multicolumn{2}{c|}{$1.082\,\pm\,0.003$}    \\
  $R^{\mathrm{NNLO}}$(\HERWIG)
& 
\multicolumn{2}{c|}{$1.501\, \pm \, 0.001$} 
& \multicolumn{2}{c|}{$1.323\, \pm\, 0.001$}      
& \multicolumn{2}{c|}{$1.146\, \pm \,  0.001$}    \\
  $R^{\mathrm{NNLO}}$(\PYTHIA)&
  \multicolumn{2}{c|}{$1.271\,\pm\,0.002$}  &
  \multicolumn{2}{c|}{$1.120\,\pm\,0.002$}     &
  \multicolumn{2}{c|}{$0.971\,\pm\,0.002$}    \\
\hline
\end{tabular}
\caption{
\label{tab:acc-cross-sections}
Comparison of the accepted cross-section obtained in fixed  order perturbation theory 
and from various event  generators. The predictions from event generators  are rescaled such
 that the predictions for the inclusive cross-section 
are in agreement with the corresponding fixed order result. }
\end{center}
\end{table}

In Table~\ref{tab:acc-cross-sections} we  present the predictions of the 
fixed order calculations  through NNLO, as  well as  the predictions of \PYTHIA,
\HERWIG\ and \mcnlo.
At this stage we do not include a simulation of hadronization and 
the underlying event, in order to make a more direct comparison with the fixed order 
predictions, which cannot take into account such effects. 
The total inclusive cross-section of \PYTHIA\ and \HERWIG\ corresponds to a pure LO computation, and that of  \mcnlo\
 is correct at NLO accuracy.  Since we  are  interested  in a comparison of the 
efficiencies and not of  absolute cross-sections, we  multiply the  results  of 
the event  generators with appropriate scaling  factors,  such that they match the 
result of the total inclusive cross-section given by our fixed order computation. Because of the different
parton distribution functions and approximation for the top loop used in the various calculations,
these scaling factors to match the leading fixed order calculations~\cite{fehip,hnnlo} 
are different from unity at the level of 5\%.

After applying the selection cuts, 
we find  a  relatively  good  agreement among the  NNLO, \mcnlo\ and \HERWIG\ 
results for the accepted cross-sections, with the \mcnlo\ and \HERWIG\ predictions rescaled to reproduce
the fixed order NNLO inclusive cross section before cuts, as explained above.
The \mcnlo~ result is  smaller than the NNLO prediction by 
$4 - 14\%$, depending on the scale choice. \HERWIG~  results differs from the NNLO prediction by $-3\%$  to $+8\%$.  On the contrary,
the accepted cross section and consequently the selection efficiency obtained with \PYTHIA\ appears 
 to be  somewhat  smaller. Depending on the scale choice the difference ranges between 12 and 21\%.

For a better understanding of the results of Table~\ref{tab:acc-cross-sections},
we now analyze  the efficiency of individual cuts, applied in turn.
The results are given in 
Table~\ref{tab:cut-efficiencies}, where the efficiencies 
due  to a  specific  cut only (after all previous cuts have been applied) are presented between parentheses. 
\begin{table}[h]
\begin{center}
\begin{tabular}{|l||c|c|c|c|}
\hline
$\accsigma/\inclsigma$ & Trigger & $+$ Jet-Veto & $+$ Isolation & All Cuts \\ 
\hline
\hline
NNLO ($\mu = \mh/2$) &  $44.7\%$ & $39.4\%$ ($88.1\%$) & $36.8\%$ $(93.4\%)$ 
& $27.8\%$ ($75.5\%$) \\  
NNLO ($\mu = 2 \; \mh$) &  $44.9\%$ & $41.8\%$ ($93.1\%$) & 
$40.7\%$ ($ 97.4\%$) & $ 31.0 \%$ ($76.2\%$) \\  
$\mcnlo$ ($\mu = \mh/2$) &  $44.4 \%$ & $ 38.1 \%$ ($85.8\%$) & $35.3\%$ ($92.5\%$) & $26.5 \%$ ($75.2\%$) \\  
$\mcnlo$ ($\mu = 2 \;\mh$) &  $44.8\%$  & $38.8\%$ ($86.7\%$)& $35.9\%$ ($92.5\%$) & $27.0\%$ ($75.2\%$) \\ 
$\HERWIG$ &  $46.7\%$ & $40.8\%$ ($87.4\%$) & $37.8\%$ ($92.7\%$) & 
$28.6\%$ ($75.7\%$)\\ 
$\PYTHIA$ & $46.6\%$ & $37.9\%$ ($81.3\%$) & $32.2\%$ ($85.0\%$)& $24.4\%$ ($75.8\%$)\\ 
\hline
\end{tabular}
\caption{
\label{tab:cut-efficiencies}
Comparison of the predicted selection efficiency after successive application 
of cuts, as obtained by fixed order calculations and event generators. 
Between parentheses we give the efficiency due to a single cut, after  all 
previous  cuts  have been applied.  
The event generator predictions correspond to the parton level only, i.e., no hadronization 
and underlying event effects are included. The first column lists the lepton selection ("trigger") efficiencies, the second (third)
columns give the results when also the jet-veto (isolation) cuts are applied in sequence and the last column lists the results after applying all remaining cuts.}
\end{center}
\end{table} 
We  observe the following: 
\begin{itemize}
\item  When only the cuts for lepton selection are applied (``Trigger''), we generally find
very good  agreement of all calculations for the corresponding 
efficiency. In detail, \mcnlo\ and NNLO yield  almost identical efficiencies, 
while \HERWIG\  and \PYTHIA\ give a  slightly higher efficiency. 

\item The veto on two or more central jets is a rather critical cut for the
achievable accuracy of the efficiency estimation.
This ``preselection'' cut  has  been used 
in the recent  Tevatron studies.  The efficiency of the  jet-veto alone, after trigger cuts, varies significantly at NNLO by about $5\%$.  As discussed in Section~\ref{sec:signal},
this is due  to the larger  scale  variation of the inclusive NNLO 
cross-section,  while at the same  time the accepted  cross-section after  the jet-veto application is  more stable. For $\mu = \mh/2$, where the fixed order 
expansion demonstrates a  faster convergence, we find that the jet-veto efficiency 
at NNLO is  in very good agreement with \mcnlo\ and \HERWIG.  \PYTHIA, which is the 
main tool used in the Tevatron analysis, predicts an efficiency which is smaller  by  
$6 \%$ for $\mu= \mh/2$.  

\item \PYTHIA\ also predicts  a  smaller efficiency by about $8\%$ for the isolation 
cut.  \HERWIG, \mcnlo\ and the NNLO calculation
are consistent, taking into account scale variations.

\item  The efficiency of the remaining  cuts is very similar in all  computations.  
After preselection,
\PYTHIA, rescaled  with an inclusive NNLO \K-factor,  
predicts between 12 and 21\% less signal events than the NNLO computation.   
\end{itemize}

The sensitivity of event generator predictions to cuts that restrict hadronic activity requires careful
investigation.  In particular, it is important to study the effect  of hadronization and the underlying event 
on the efficiency. We have performed such an analysis in the case of \mcnlo, where hadronization
effects are modeled by \HERWIG\ and the underlying event is simulated by interfacing to the \JIMMY\ package
 \cite{Butterworth:1996zw}, and also for \PYTHIA, where the hadronization and underlying event modelling is
inbuilt.  The results are given in Table \ref{tab:cut-efficiencies-mcnlo}.  We  find only minor differences at the
1\% level, which leads us to the conclusion that the differences observed with \PYTHIA\ are rather related to its
matrix element and parton shower implementation.    
\begin{table}[h]
\begin{center}
\begin{tabular}{|l||c|c|c|c|}
\hline
$\accsigma/\inclsigma$ & Trigger & $+$ Jet-Veto & $+$ Isolation & All Cuts \\ 
\hline
\hline
$\mcnlo$ (Parton) &  $44.4 \%$ & $ 38.1\%$ & $35.3\%$ & $26.5 \%$ \\
$\mcnlo$ (Hadron) &  $44.6 \%$ & $ 39.6 \%$ & $37.4\%$ & $28.3 \%$ \\
$\mcnlo$ (Had + UE) &  $44.6\%$  & $39.4\%$ & $36.5 \%$ & $27.6\%$  \\
\hline
\PYTHIA\ (Parton) & $46.6\%$ & $37.9\%$ & $32.2\%$ & $24.4\%$ \\
\PYTHIA\ (Had + UE) & $44.5\%$ & $37.2\%$ & $31.3\%$ & $23.8\%$ \\ 
\hline
\end{tabular}
\caption{ 
  \label{tab:cut-efficiencies-mcnlo} Selection efficiency obtained with \mcnlo\ ($\mu=m_H/2$) and \PYTHIA\ after
     successive application of cuts. The first row gives the results for the purely partonic \mcnlo\  simulation, in the
     second it is complemented by the cluster hadronization model as implemented in \HERWIG, and in the third an
      underlying event model (\JIMMY) is added to the simulation.  For \PYTHIA\ the inbuilt hadronization and    
      underlying event models are applied simultaneously.
}
\end{center}
\end{table}

\begin{figure}[th]
  \begin{center}
    \includegraphics[width=0.48\textwidth]{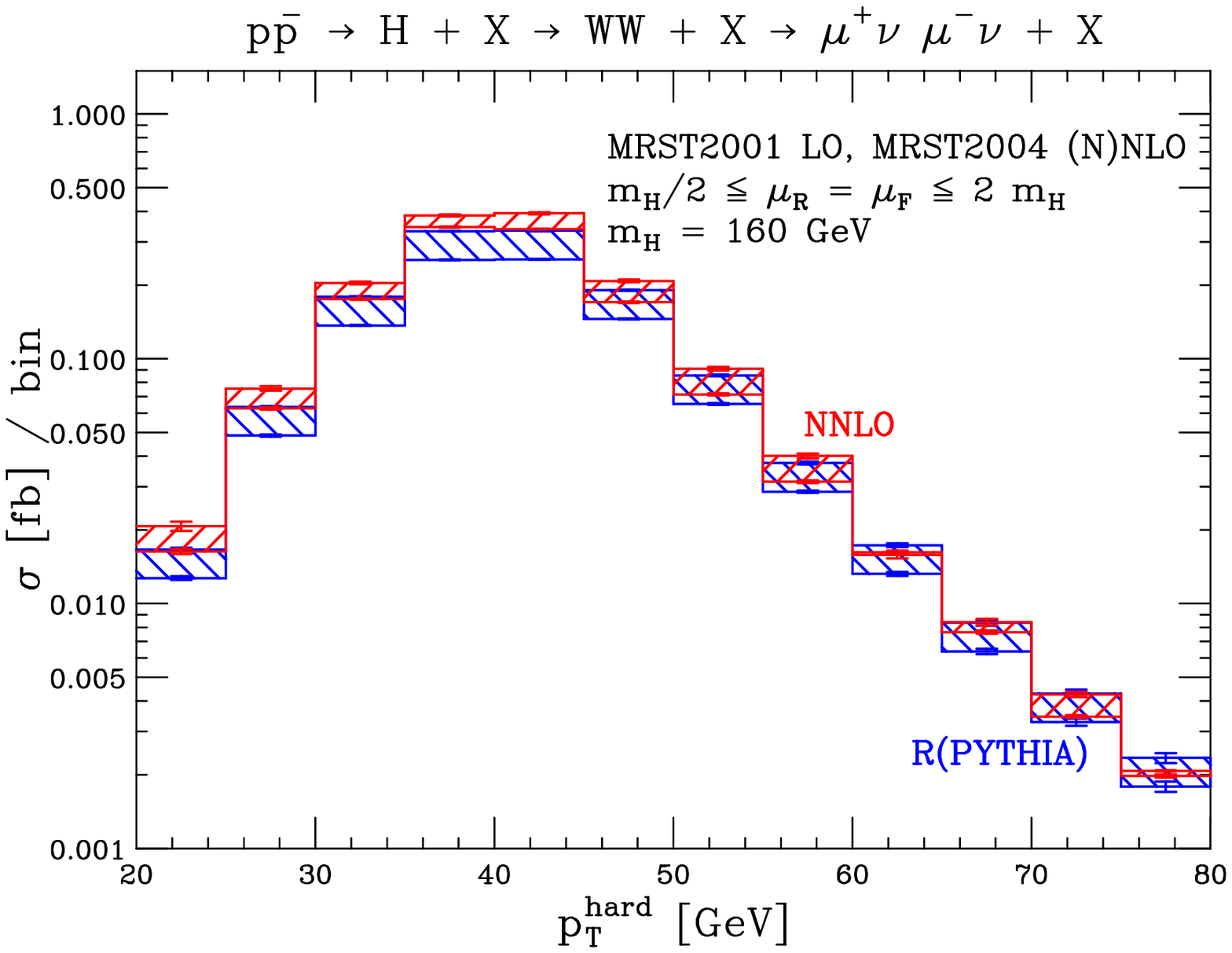}
    \includegraphics[width=0.48\textwidth]{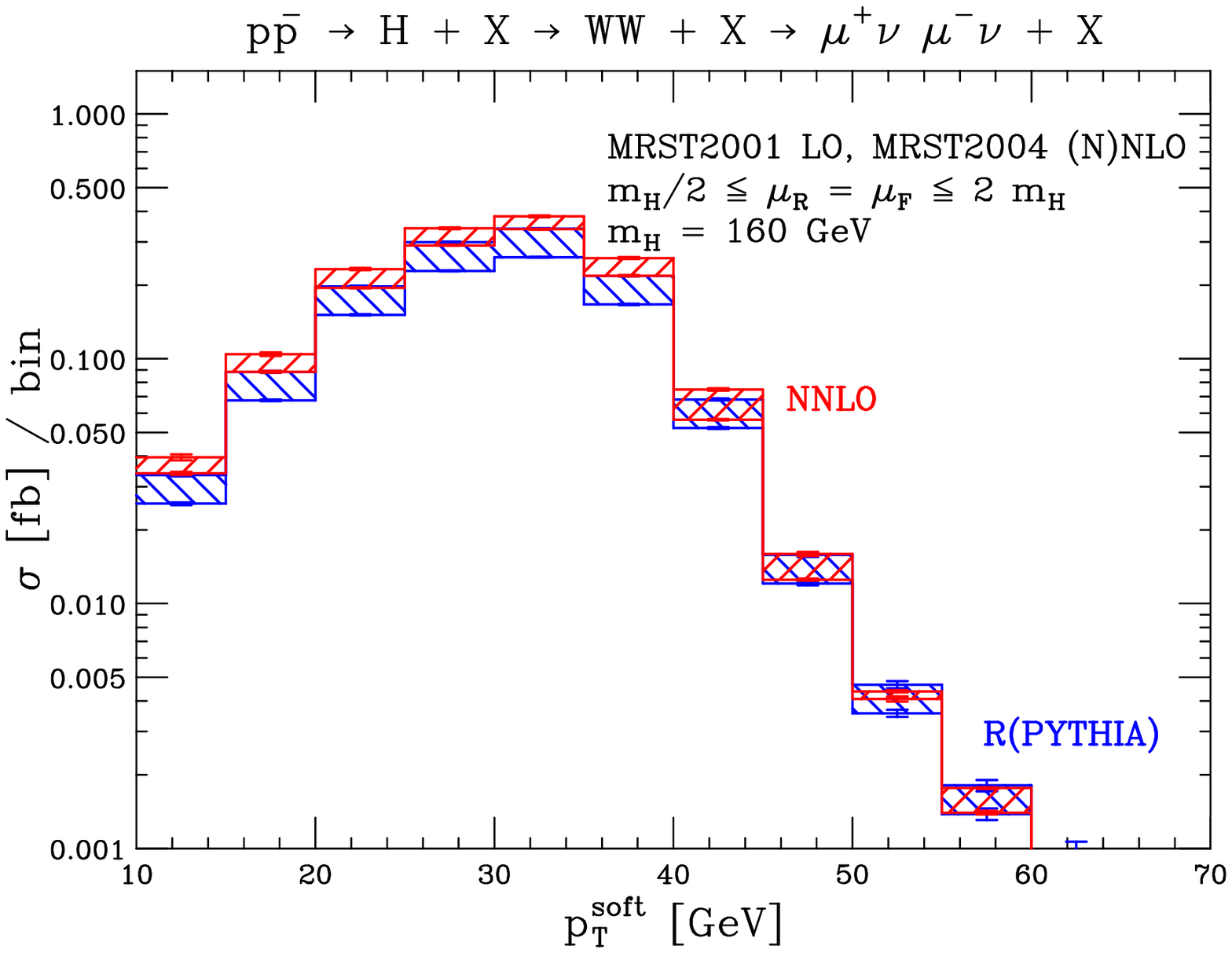}\\
    \includegraphics[width=0.48\textwidth]{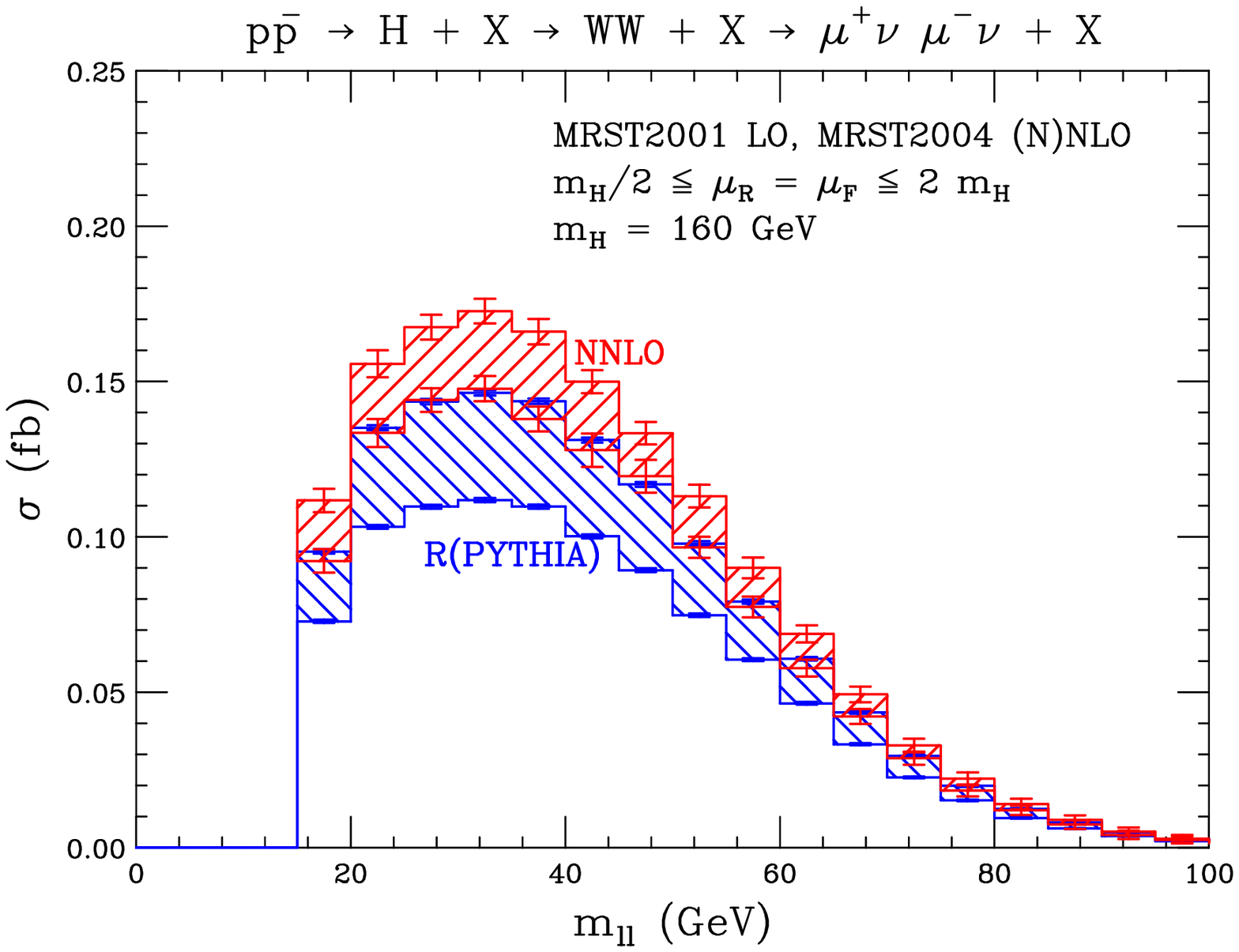}
    \includegraphics[width=0.48\textwidth]{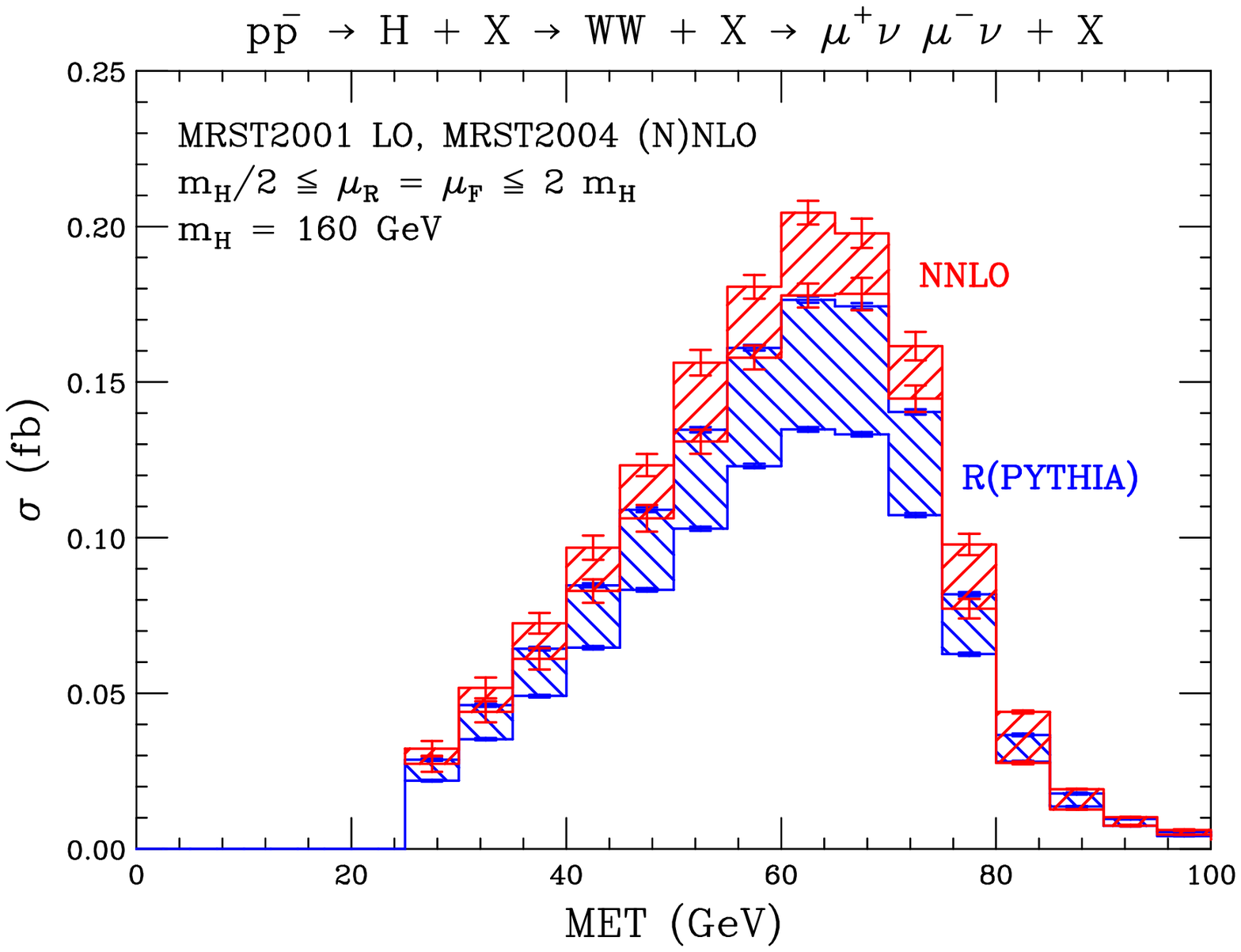}\\
    \includegraphics[width=0.48\textwidth]{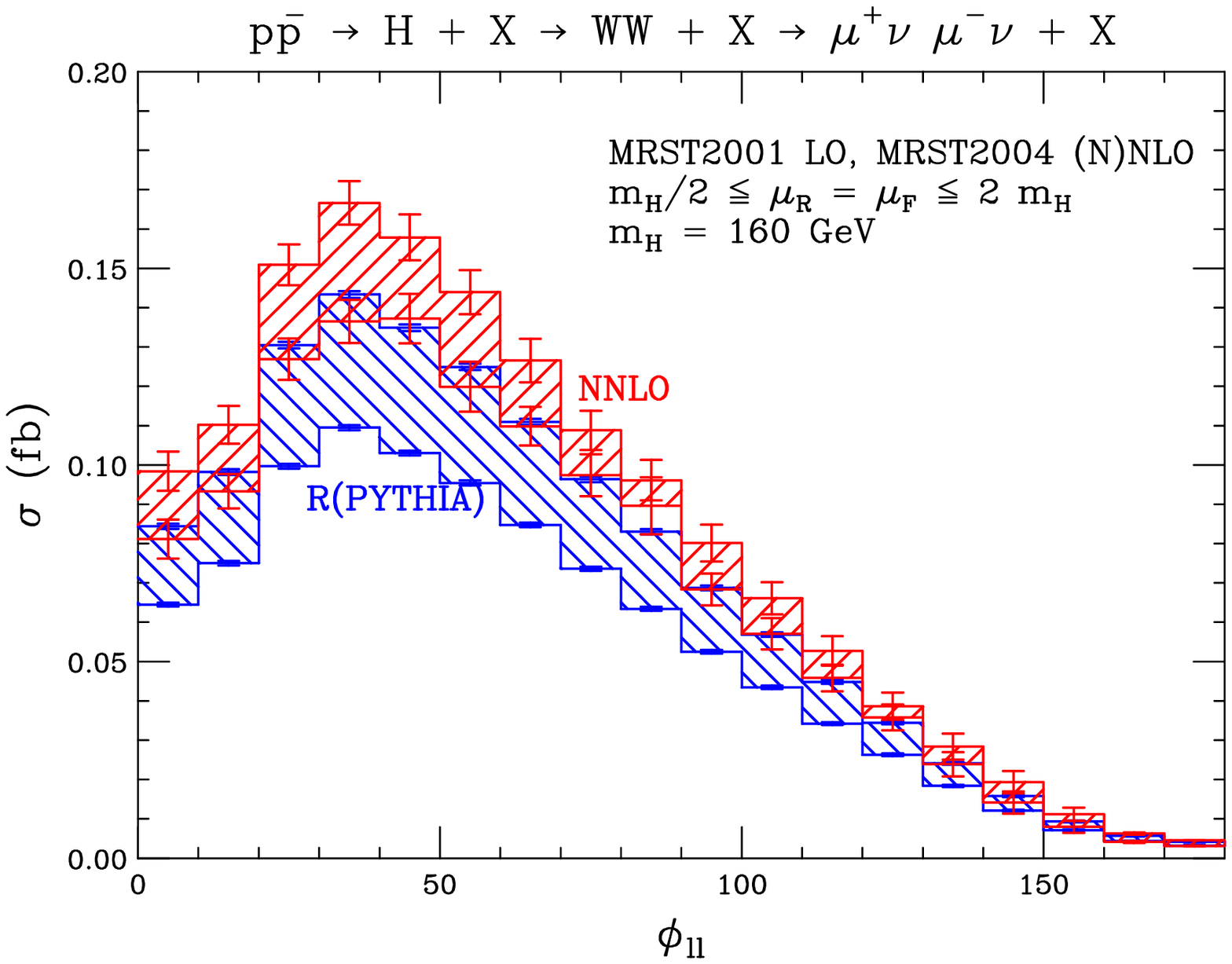}
  \end{center}
  \caption{\label{fig:pythia_distr}
  Kinematic distributions obtained at NNLO in perturbative QCD and with \PYTHIA. The \PYTHIA\ results are rescaled by
   an inclusive \K-factor in order to reproduce the inclusive cross section at NNLO. Shown are the transverse
    momentum of the leading ($\pthard$) and trailing lepton ($\ptsoft$), the invariant mass of the two leptons ($m_{ll}$),
    the transverse missing energy (MET) and the azimuthal separation of the two charged leptons in the transverse plane ($\phill$).
}
 \end{figure}

\begin{figure}[th]
  \begin{center}
    \includegraphics[width=0.48\textwidth]{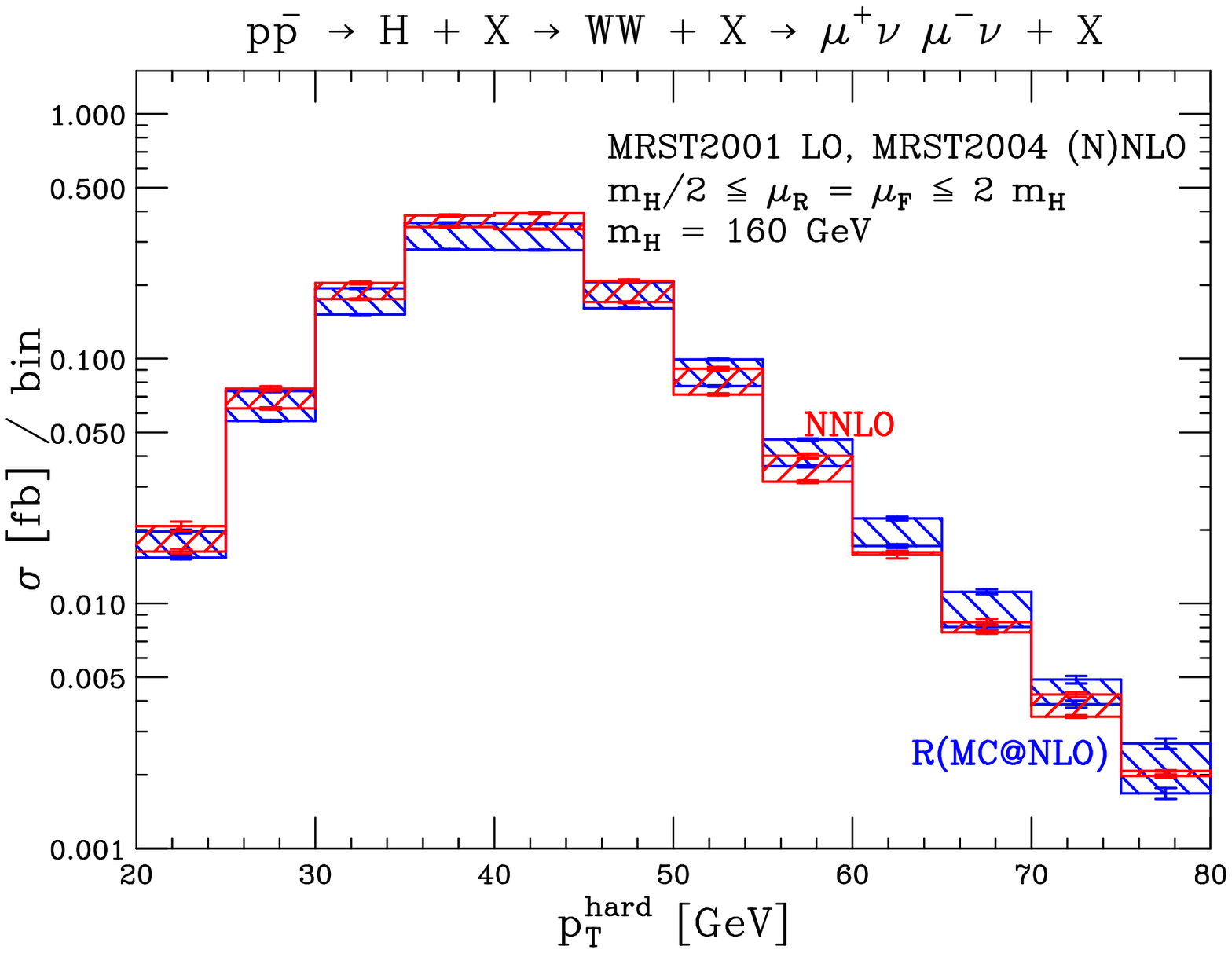}
    \includegraphics[width=0.48\textwidth]{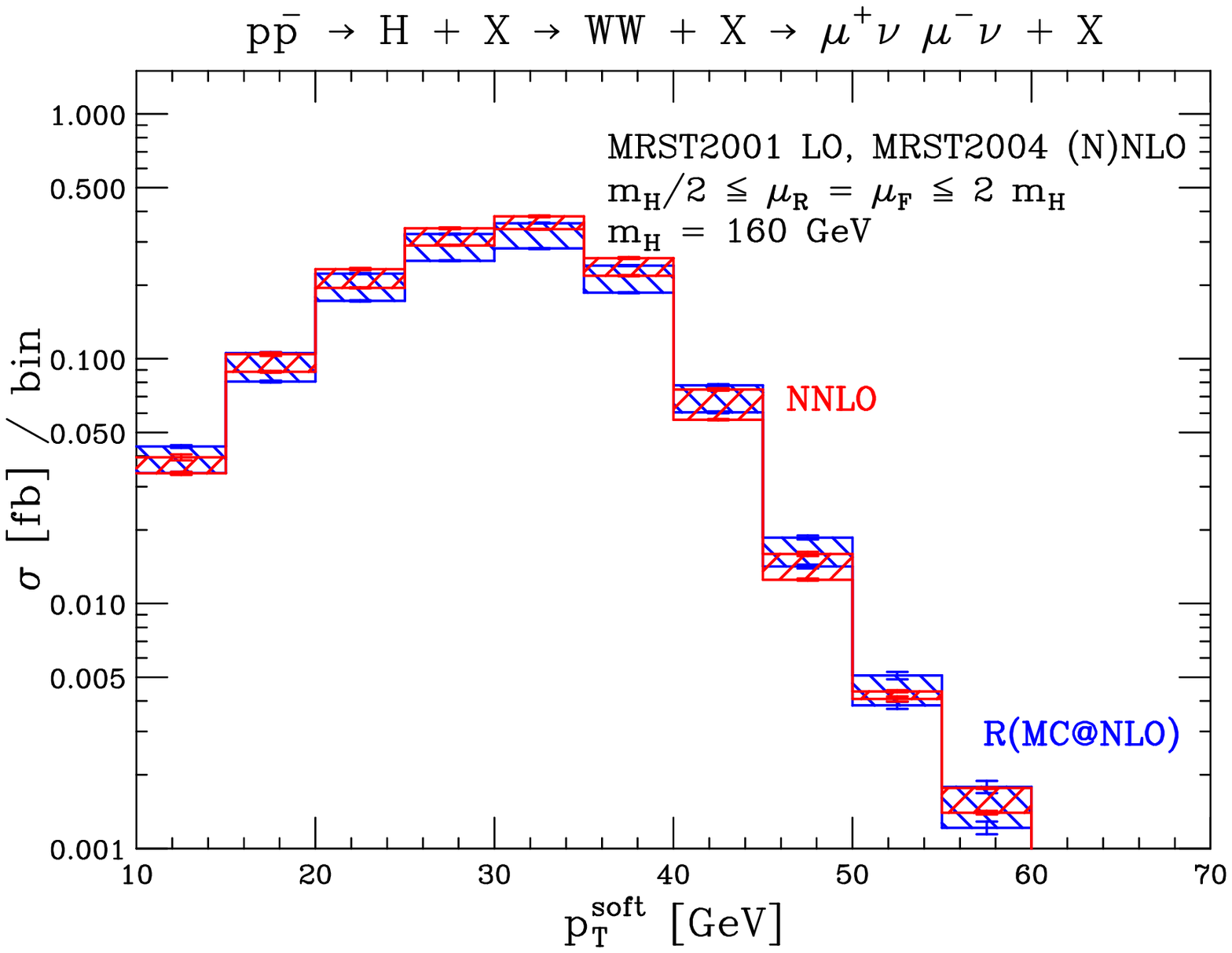}\\
    \includegraphics[width=0.48\textwidth]{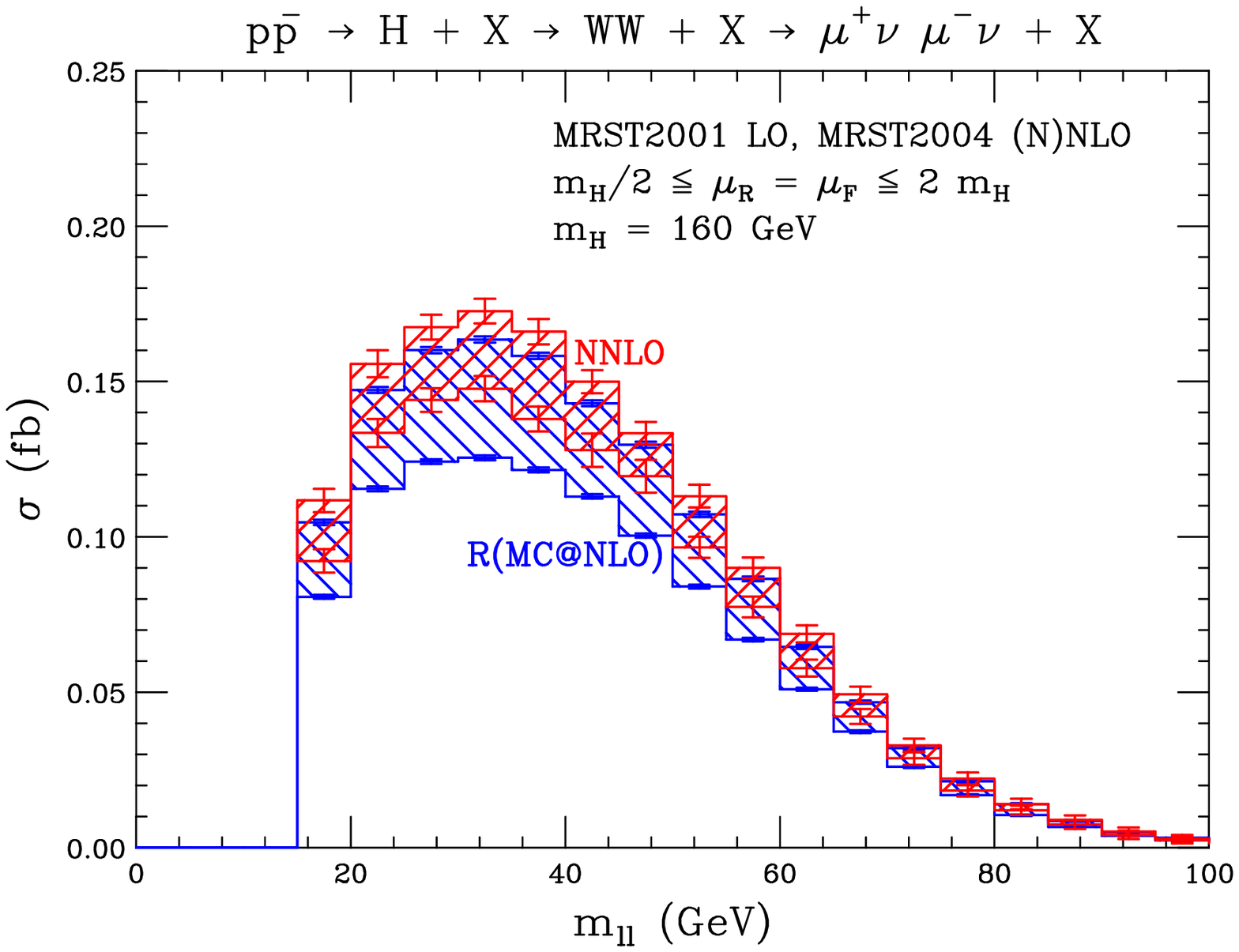}
    \includegraphics[width=0.48\textwidth]{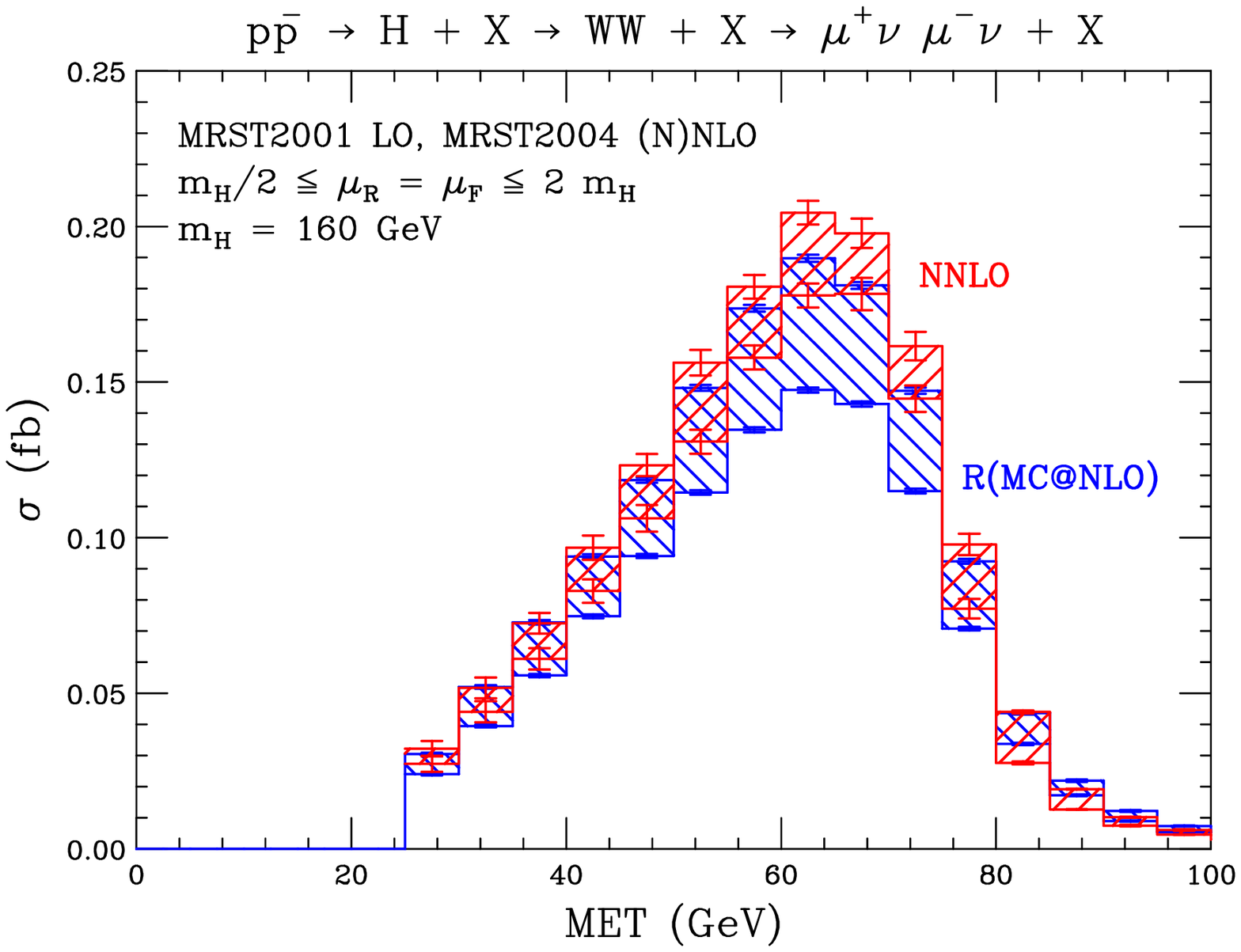}\\
    \includegraphics[width=0.48\textwidth]{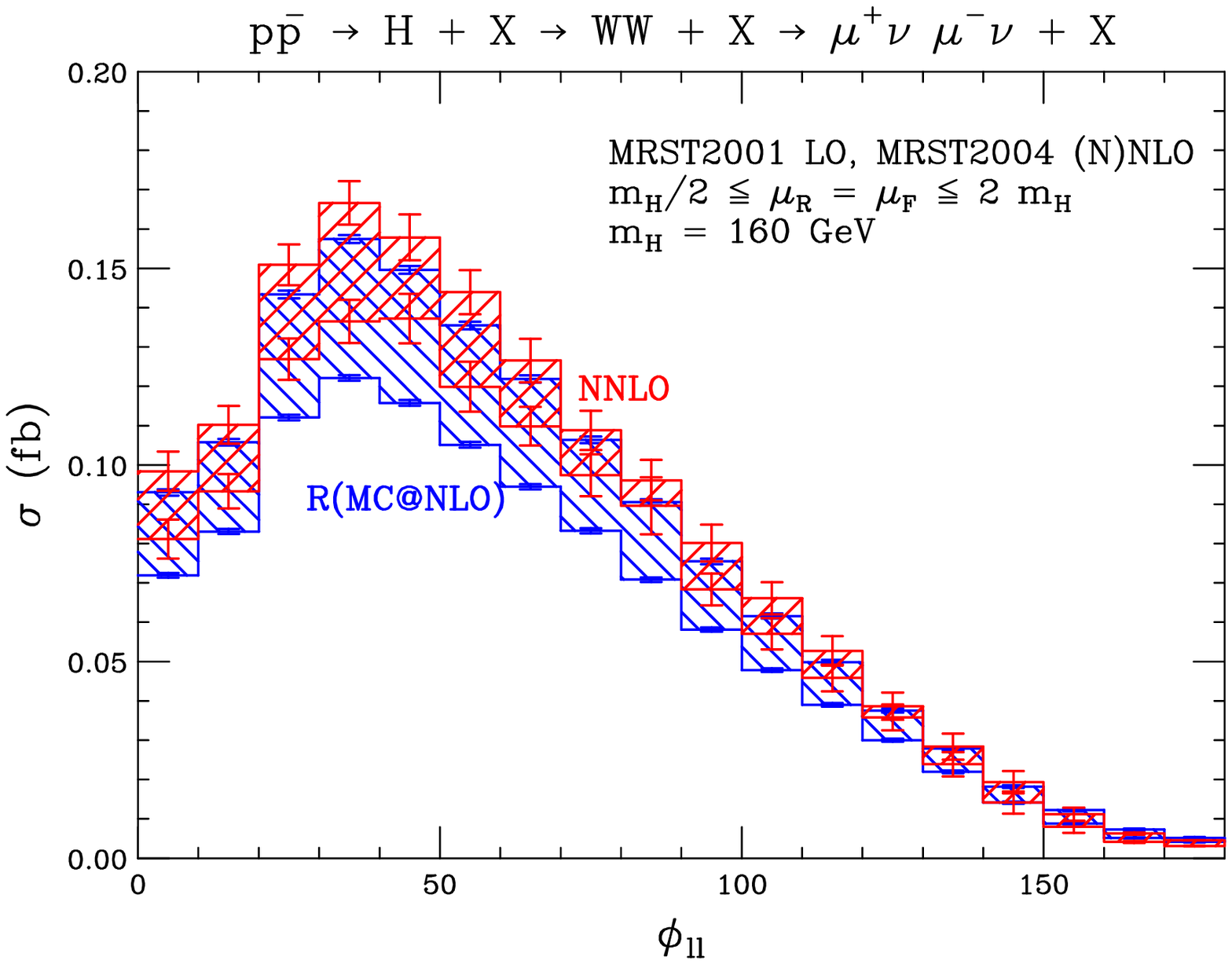}
  \end{center}
  \caption{\label{fig:mcatnlo_distr}
  Kinematic distributions obtained at NNLO in perturbative QCD and with \mcnlo. The \mcnlo\ results are rescaled by
   an inclusive \K-factor in order to reproduce the inclusive cross section at NNLO. Shown are the transverse
    momentum of the leading ($\pthard$) and trailing lepton ($\ptsoft$), the invariant mass of the two leptons ($m_{ll}$),
    the transverse missing energy (MET) and the azimuthal separation of the two charged leptons in the transverse plane ($\phill$).
}
\end{figure}

\section{Artificial Neural Network}
\label{sec:ANN}

The current experimental analysis at the Tevatron attempts to distinguish  a very small 
number  of events  from a considerably larger background. For such a task, the use  of  advanced  
statistical methods is necessary. An integral part of  the experimental 
studies  are  distributions of discrimination variables, defined via 
artificial neural networks (ANN). It is  clear that such techniques 
will become an indispensable tool in many future studies  at the Tevatron and 
at  the LHC, optimizing the sensitivity of the experiments.  

To the best of our  knowledge, so far there has  been no study  of  how the distributions 
of ANN outputs are modified  at  higher orders in  perturbation theory. 
Here we present for the first  time an ANN output distribution, computed at fixed  order in perturbation 
theory, beyond the leading order.

In order to study these higher-order effects on the outcome variable,
we have built an ANN with the tool TMVA v.~3.9.2.~\cite{tmva} based on
the Data Analysis Framework ROOT v.~5.21.02~\cite{root}. In the
construction (the so-called training) of the ANN the user has to provide a
set of signal and background events, as well as a list of
input variables. In our study we use the variables defined in
Section~\ref{sec:kinematics_nnlo} as input variables. Based on the techniques of
Multilayer Perceptrons, the ANN then builds an output variable, which is basically a
non-linear function of the input variables. Since our study is based
on Monte Carlo truth information, we restrict the set of background 
to processes that have the same final state signature as the signal,
i.e.\ to continuum WW background and top-pair production. For both of these
processes, as well as for the signal process, we generate a large
enough event sample ($\sim 50000$ events after pre-selection) with the LO parton-shower
Monte Carlo \PYTHIA8~\cite{pythia}, and use half of the samples to train the ANN. 
The other half is used in the so-called testing step.

\begin{figure}[h]
  \begin{center}
    \includegraphics[width=0.48\textwidth]{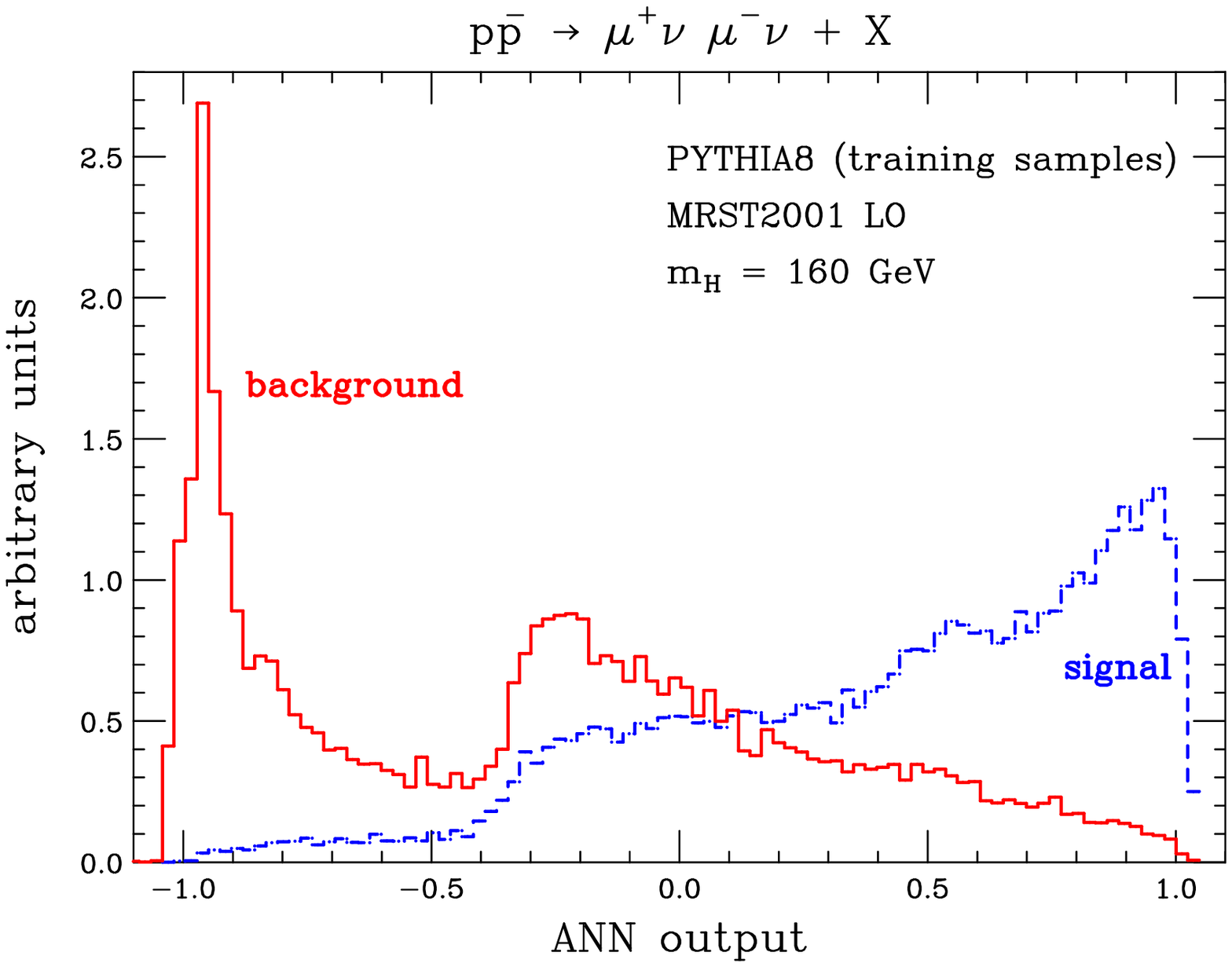}
    \includegraphics[width=0.48\textwidth]{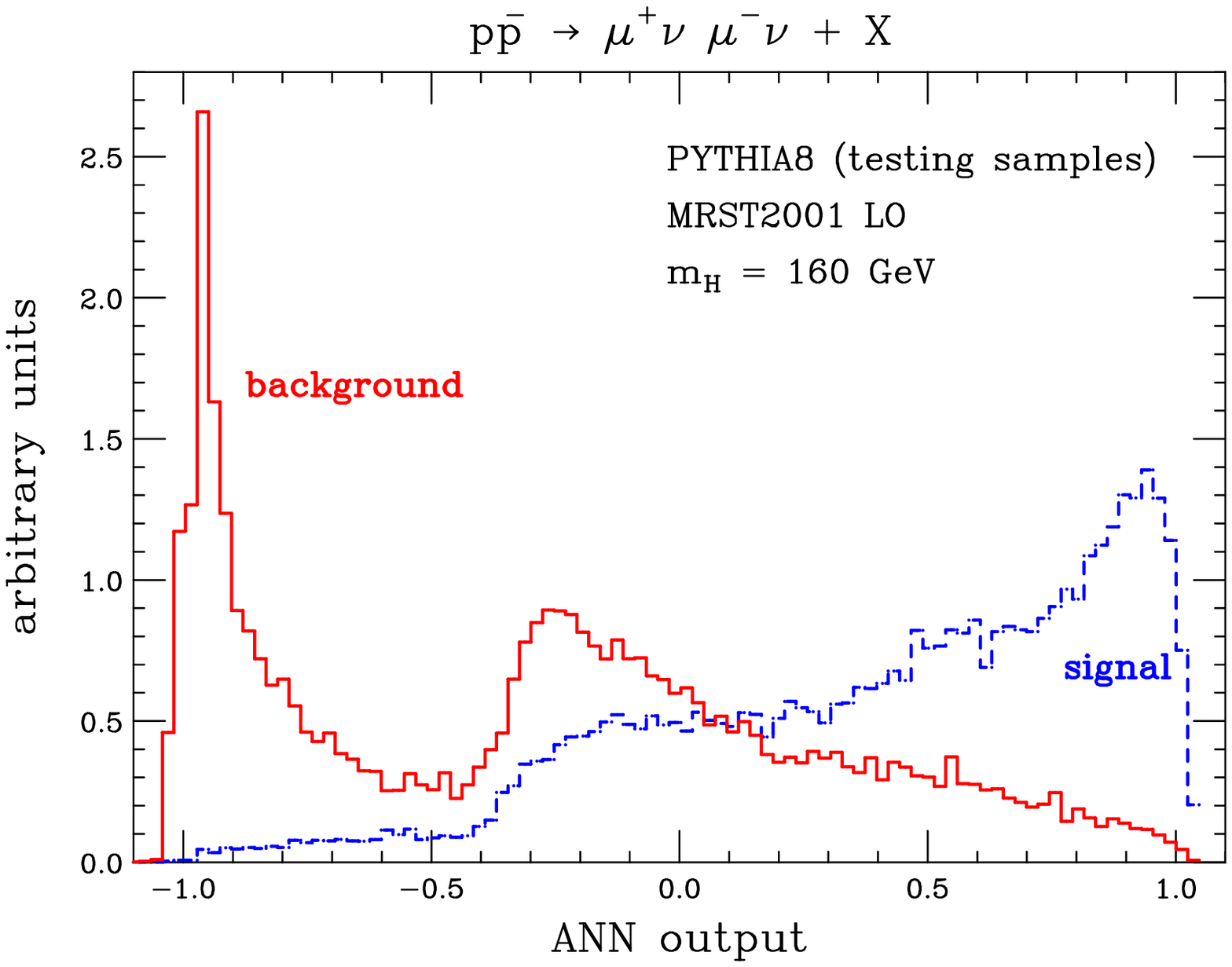}
  \end{center}
  \caption{\label{fig:ANNtraining}
Distribution of the ANN output variable coming from the ANN
construction procedure. On the left, the distribution for the training
sample, on the right for the testing sample.}
\end{figure}

Fig.~\ref{fig:ANNtraining} shows the ANN output for the signal and background samples,
for the training (left) and the testing (right) samples.
The comparison of the training and testing distributions serves as
a verification that the ANN has not been over-trained, i.e.\ tuned to
statistical effects in the training sample. The discrimination power
of the ANN variable can be seen clearly. While the background
distribution peaks at low values,\footnote{The background peaks around $-1.0$ and $-0.3$
arise mainly from $t\bar t$ and $WW$ respectively.}
the signal events populate the high value range. Distributions like this 
can serve to distinguish background from possible signal events, either
by cutting on the ANN output, or by using the shape of the
distribution to decide whether the observed event set consists of
background only or background and signal events. 

\begin{figure}[h]
  \begin{center}
    \includegraphics[width=0.78\textwidth]{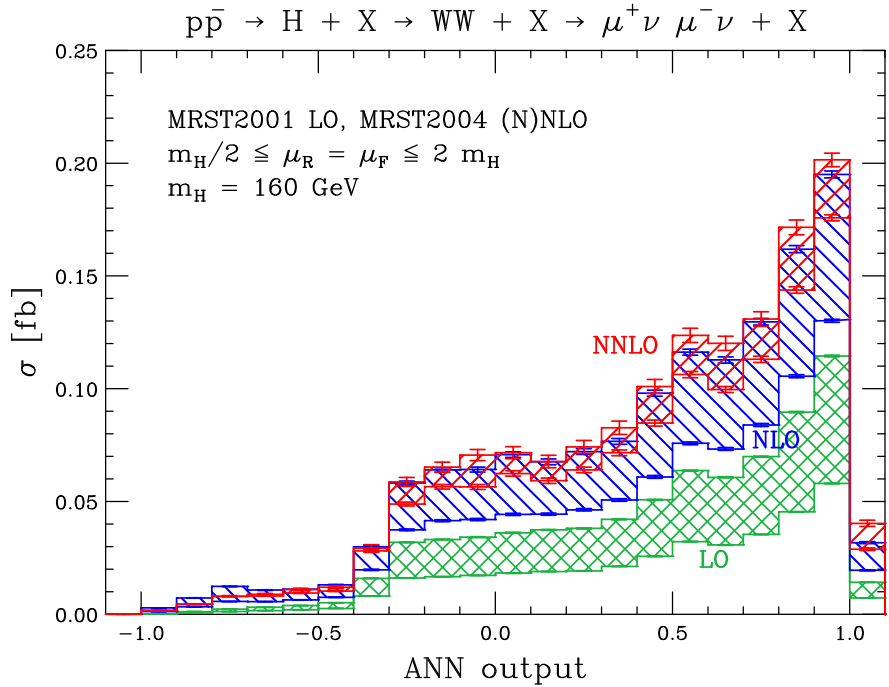}
  \end{center}
  \caption{\label{fig:ANN_nnlo}
Distribution of the ANN output variable, computed at LO, NLO and NNLO in perturbation theory. The bands indicate the
scale uncertainties.}
\end{figure}

In Fig.~\ref{fig:ANN_nnlo} we present the distribution of  the ANN output for the
signal in fixed-order perturbation theory, computed  
at LO, NLO and  NNLO.  We find  significant radiative corrections, which however are 
consistent in magnitude with those for the accepted  cross section
after preselection cuts.

\begin{figure}[th]
  \begin{center}
    \includegraphics[width=0.78\textwidth]{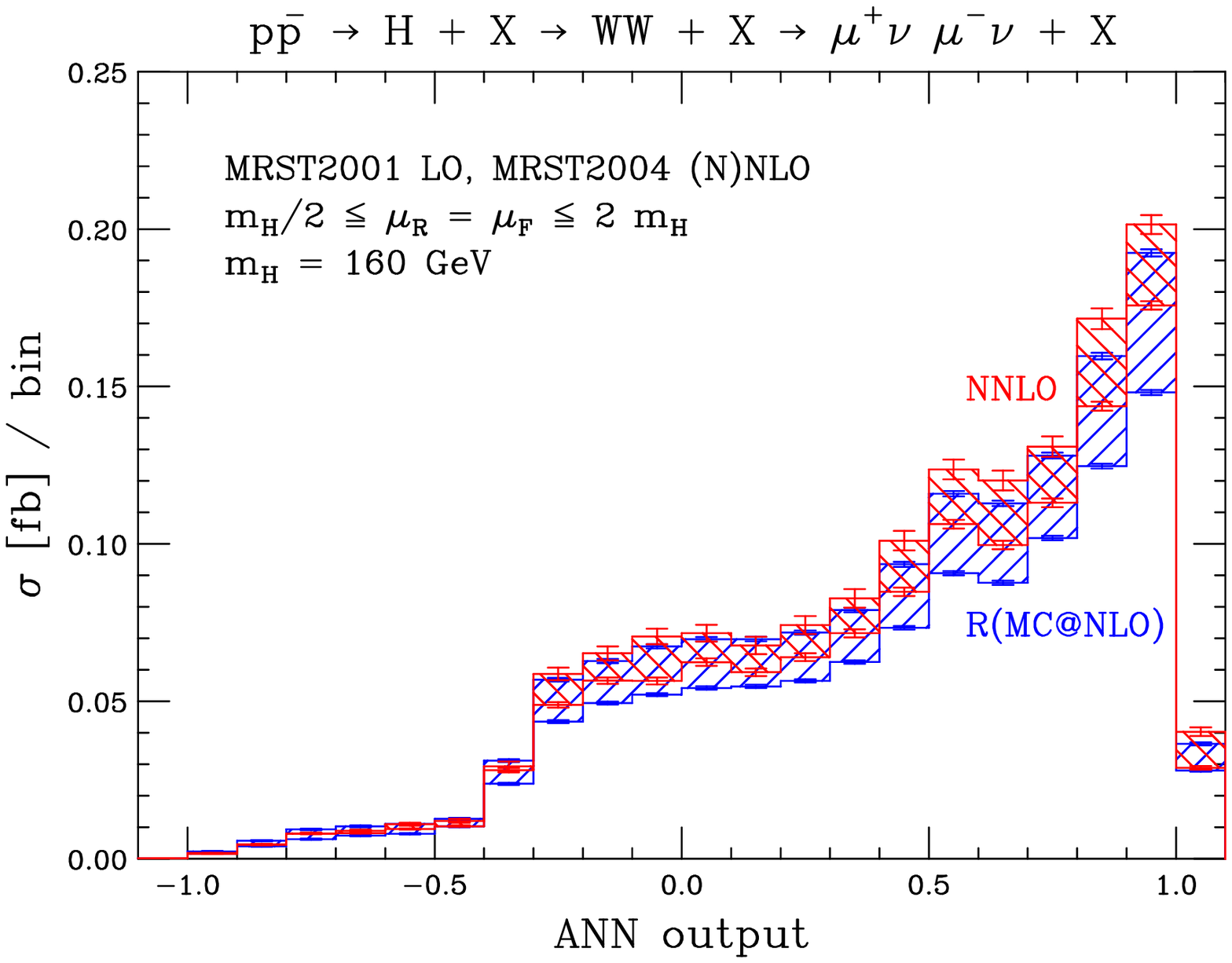}
  \end{center}
  \caption{\label{fig:mcatnlo_ANN}
    Distribution of the ANN output variable, computed at NNLO in perturbation theory and with
  \mcnlo. The \mcnlo\ result is rescaled by
   an inclusive \K-factor in order to reproduce the inclusive cross section at NNLO. 
   The bands indicate the scale uncertainties.}
\end{figure}

In Fig.~\ref{fig:mcatnlo_ANN} we compare the ANN distribution obtained at NNLO and
with \mcnlo. 
Again we  find a very good agreement  between the fixed order calculation and 
the \mcnlo\ prediction, when the latter is rescaled with a \K-factor in order to reproduce the 
total inclusive cross-section. 

\begin{figure}[th]
  \begin{center}
    \includegraphics[width=0.78\textwidth]{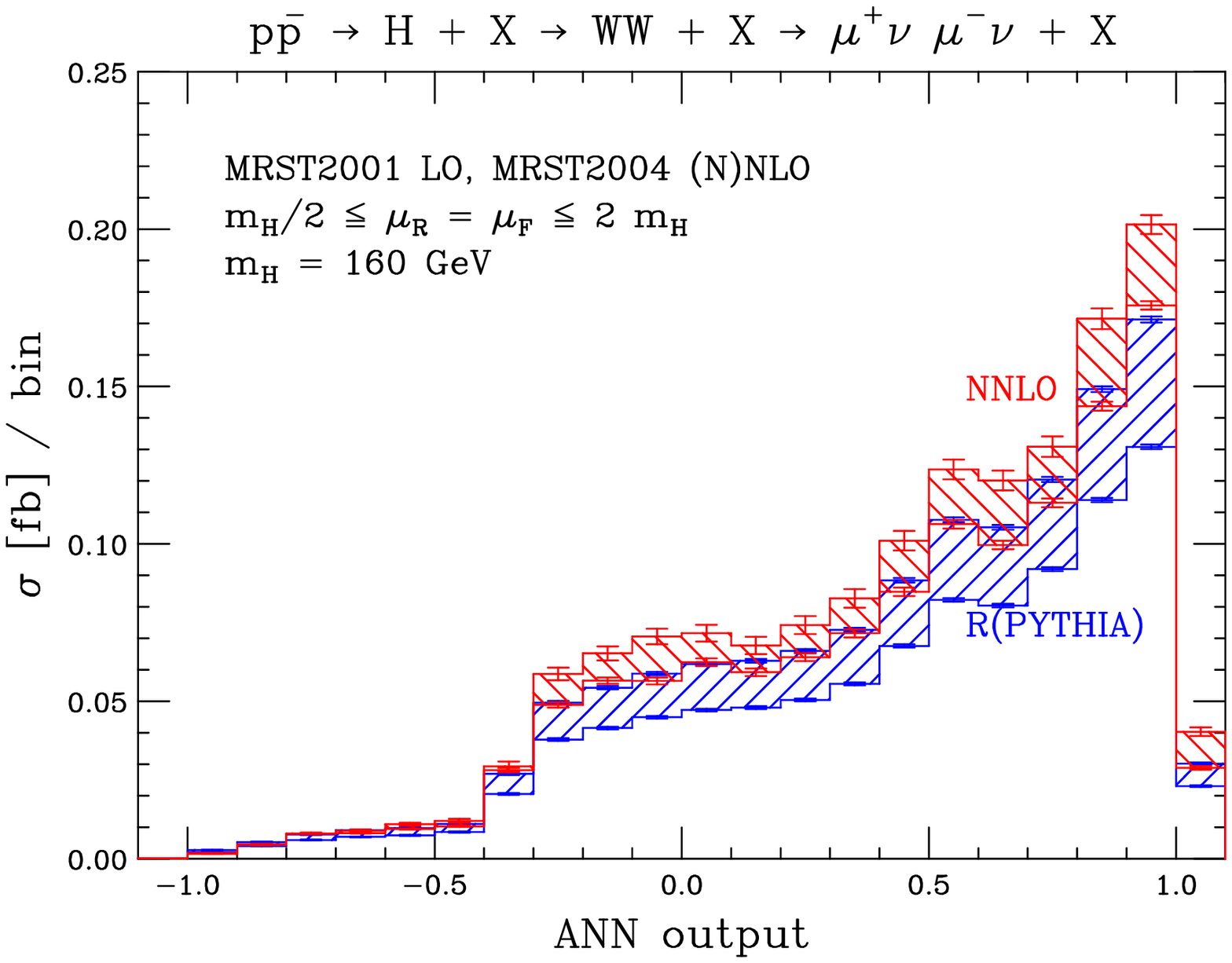}
  \end{center}
  \caption{\label{fig:pythia_ANN}
  Distribution of the ANN output variable, computed at NNLO in perturbation theory and with
  \PYTHIA. The \PYTHIA\ result is rescaled by
   an inclusive \K-factor in order to reproduce the inclusive cross section at NNLO. 
   The bands indicate the scale uncertainties.}
\end{figure}

Finally, in Fig.~\ref{fig:pythia_ANN} we compare the ANN distribution obtained at NNLO QCD and
with \PYTHIA. We see that \PYTHIA, even after rescaling with an inclusive  \K-factor, yields predictions 
which are smaller by 12-20\%, depending on the chosen bin.   This difference can be traced back to
the difference in efficiency already observed at the level of the 
selection cuts placed on the kinematic input distributions.

Note that we have not included any hadronic variable as an input to 
the ANN. It is clear that stable perturbative patterns are obtained as 
long as we  apply cuts on ``leptonic'' variables only.  However, adding a hadronic 
variable to the list of ANN inputs could produce results that are very sensitive to the details of the  modeling of the hadronic activity in the event generators used for the training of the network. 

\section{Conclusions}
\label{sec:TheEnd}

In this paper we have studied higher-order QCD effects in the search for a Higgs boson of mass $\mh=160$ GeV at the Tevatron. We have considered a definite set of preselection cuts that we believe capture the essential features
of the CDF and D\O~analyses. We
have studied the impact of higher order corrections on a set of kinematical distributions of the final state leptons.
We have then compared these distributions, computed up to NNLO in QCD perturbation theory, to those obtained
with the \PYTHIA, \HERWIG\ and \mcnlo\ event generators.  The comparison of distributions
does not show significant differences, and this is confirmed by a more sophisticated
analysis we have performed based on the training of our own ANN.

For the ANN analysis we used only leptonic input variables, in order to reduce sensitivity to the modelling of hadronic
activity and to allow perturbative evaluation of the output distribution.  These features are also necessary for the
reliable estimation of theoretical uncertainties in experimental ANN analyses.

We have also compared the efficiency of the experimental cuts obtained in NNLO QCD to those obtained with \PYTHIA, \HERWIG\ and \mcnlo. The efficiencies obtained
with \HERWIG\ and \mcnlo\ are consistent with that obtained at NNLO.
The \mcnlo\ acceptance is slightly smaller than the NNLO acceptance, by 
$4 - 14\%$, while the acceptance of  \HERWIG\  differs from the NNLO prediction by $-3\%$  to $+8\%$. 
In contrast, we  find that the acceptance  computed with  \PYTHIA\  is between $12\%$ and $21\%$ smaller  than the NNLO acceptance, depending on the choice of the factorization and renormalization scale.
This result is not significantly altered by hadronization and underlying event and appears instead
to be related to the matrix element and parton shower implementation in \PYTHIA\ itself.    
Since the Tevatron analyses are based on \PYTHIA, we believe that this effect could be important
and requires a more detailed investigation within the framework of the full experimental analysis.

Relevant to the experimental analysis, we have remarked that the combination in quadrature of the theoretical
errors due to the parton distributions and scale variations in Refs.~\cite{CDFnote,DOnote} 
implies that the theoretical uncertainty on the total cross section used there is likely to be underestimated.
We also pointed out that a reweighting of parton
shower Monte-Carlos to match the fixed-order Higgs $p_T$ distribution is not appropriate for
events with a low Higgs $p_T$ value. 
Finally, we have demonstrated that a reliable estimation of the theoretical uncertainty
for Higgs signal cross-sections with defined jet multiplicities requires dedicated fixed order
computations for each multiplicity. The theoretical uncertainty in each jet-bin is different
from the theoretical uncertainty of the total cross-section.

\section*{Acknowledgments}

We thank Stefan Bucherer, Frank Petriello, Uli Haisch, Zoltan Kunszt and Giulia Zanderighi for usefull discussions.


\bibliographystyle{JHEP}

\end{document}